\documentclass[twocolumn]{IEEEtran}

\usepackage{amsmath}

\DeclareMathOperator*{\argmin}{arg\,min}

\usepackage{amsfonts,amssymb,amsbsy}
\usepackage{mathrsfs}
\usepackage{hyperref}
\usepackage[dvipsnames]{xcolor}
\usepackage{pdfpages}

\usepackage{graphicx}
\usepackage{subcaption} 
\usepackage{pgfplots}
\usepackage[mode=tex]{standalone}

\usepackage{cleveref}

\usepackage{times,epsfig}
\usepackage{cite}
\usepackage{cases}
\usepackage{setspace}
\usepackage{algorithmic}
\usepackage{algorithm}

\usepackage{balance}

\newcommand{\numNodes}{\mathrm{N}}
\newcommand{\numNonzero}{\mathrm{S}}

\newcommand{\numSources}{\mathrm{R}}
\newcommand{\numTaps}{\mathrm{L}}
\newcommand{\nodeIndex}{\mathrm{n}}
\newcommand{\sourceIndex}{\mathrm{i}}
\newcommand{\otherSourceIndex}{\mathrm{j}}
\newcommand{\yetOtherSourceIndex}{\mathrm{k}}
\newcommand{\tapIndex}{\mathrm{l}}
\newcommand{\numSamples}{\mathrm{M}}

\DeclareMathOperator{\diag}{diag}

\DeclareMathOperator{\vect}{vec}

\newcommand{\decoupledmodelop}{\mathcal{A}}
\newcommand{\modelop}{\mathcal{M}}
\newcommand{\project}{\mathcal{P}}

\newcommand{\mZ}{\mathcal{Z}}

\newcommand{\mS}{\Omega}
\newcommand{\mT}{\mathrm{T}}

\newcommand{\numPartionElems}{\text{P}}
\newcommand{\partion}{\Gamma}
\newcommand{\partionIndex}{\text{p}}
\newcommand{\partionSize}{\mathrm{Q}}
\newcommand{\partmat}{\mathbf{D}}
\newcommand{\parimat}{\mathbf{F}}

\newcommand{\ba}{\mathbf{a}}
\newcommand{\bb}{\mathbf{b}}
\newcommand{\bh}{\mathbf{h}}
\newcommand{\bl}{\mathbf{l}}
\newcommand{\bq}{\mathbf{q}}
\newcommand{\br}{\mathbf{r}}
\newcommand{\bs}{\mathbf{s}}
\newcommand{\bu}{\mathbf{u}}

\newcommand{\bx}{\mathbf{x}}
\newcommand{\by}{\mathbf{y}}
\newcommand{\bz}{\mathbf{z}}

\newcommand{\bA}{\mathbf{A}}
\newcommand{\bB}{\mathbf{B}}
\newcommand{\bC}{\mathbf{C}}
\newcommand{\bH}{\mathbf{H}}
\newcommand{\bI}{\mathbf{I}}
\newcommand{\bL}{\mathbf{L}}
\newcommand{\bM}{\mathbf{M}}
\newcommand{\bP}{\mathbf{P}}
\newcommand{\bR}{\mathbf{R}}
\newcommand{\bS}{\mathbf{S}}
\newcommand{\bT}{\mathbf{T}}
\newcommand{\bU}{\mathbf{U}}
\newcommand{\bV}{\mathbf{V}}
\newcommand{\bW}{\mathbf{W}}
\newcommand{\bX}{\mathbf{X}}
\newcommand{\bY}{\mathbf{Y}}
\newcommand{\bZ}{\mathbf{Z}}

\newcommand{\bzero}{\mathbf{0}}

\newcommand{\blambda}{\boldsymbol{\lambda}}
\newcommand{\bnu}{\boldsymbol{\nu}}
\newcommand{\bpsi}{\boldsymbol{\psi}}

\newcommand{\bLambda}{\boldsymbol{\Lambda}}
\newcommand{\bPsi}{\mathbf{\Psi}}
\newcommand{\bSigma}{\mathbf{\Sigma}}

\newcommand{\transpose}{^\top}
\newcommand{\hermitian}{^\text{H}}

\newcommand{\evn}{\mathbb{E}}
\newcommand{\ev}[1]{\evn\left(#1\right)}

\newcommand{\norm}[1]{\left\|{#1}\right\|}

\newcommand{\tF}{\text{F}}
\newcommand{\tK}{\text{K}}

\newcommand{\boundDualCertificateSecondCondition}{\partionSize \geq \mathcal{C}_{\alpha+\log\numSources} \numSources \textcolor{blue}{\max\{\mu_{\max}^2 \numTaps_\sourceIndex \rho_{\max}, \mu_h^2 (\rho_{\max} +1)^2 \numNonzero_\sourceIndex\}} \log^2\numNodes}

\newcommand{\estimate}[1]{\breve{#1}}

\usepackage{theorem}
\newtheorem{mytheorem}{\bf Theorem}

\newtheorem{mylemma}{\bf Lemma}
\newtheorem{myproposition}{\bf Proposition}
\newtheorem{remark}{\bf Remark}

\newcommand*{\QEDA}{\hfill\ensuremath{\blacksquare}}%

\newenvironment{myproof}[1][$\!\!$]{{\noindent\emph {Proof #1:} }}{\hfill$\QEDA$}

\title{Blind Demixing of Diffused Graph Signals}

\author{\IEEEauthorblockN{Fernando J. Iglesias Garcia, Santiago Segarra,~\IEEEmembership{Senior Member,~IEEE},\\Antonio G. Marques,~\IEEEmembership{Senior Member,~IEEE}}
\thanks{Work in this paper is supported by Spanish Fed. Grants KLINILYCS TEC2016-75361-R, SPGraph PID2019-105032GB-I00, by the Grants F661-MAPPING-UCI and F663-AAGNCS funded
	by the Comunidad de Madrid (CAM) and 
	King Juan Carlos University
	(URJC), and by the USA NSF under award CCF-2008555. Fernando J. Iglesias Garcia and Antonio G. Marques are with the Dept. of Signal Theory and Comms., King Juan Carlos Univ. Santiago Segarra is with the Dept. of Electrical and Computer Eng., Rice University.  Emails: fj.iglesias@alumnos.urjc.es, segarra@rice.edu, antonio.garcia.marques@urjc.es. Preliminary results were published in the conference paper~\cite{iglesias2018demixing}.}}

\begin{document}
\maketitle

\markboth{IEEE TRANSACTIONS ON SIGNAL PROCESSING (FIRST SUBMISSION 23 December 2020, CURRENT VERSION April 2023)}
{IGLESIAS \MakeLowercase{\textit{et al.}}: Blind Demixing of Diffused Graph Signals}

\begin{abstract}
Using graphs to model irregular information domains is an effective approach to deal with some of the intricacies of contemporary (network) data. A key aspect is how the data, represented as graph signals, depend on the topology of the graph. Widely-used approaches assume that the observed signals can be viewed as outputs of graph filters (i.e., polynomials of a matrix representation of the graph) whose inputs have a particular structure. Diffused graph signals, which correspond to an originally sparse (node-localized) signal percolated through the graph via filtering, fall into this class. In that context, this paper deals with the problem of jointly identifying graph filters and separating their (sparse) input signals from a mixture of diffused graph signals, thus generalizing to the graph signal processing framework the classical blind demixing (blind source separation) of temporal and spatial signals. We first consider the scenario where the supporting graphs are different across the signals, providing a theorem for demixing feasibility along with probabilistic bounds on successful recovery. 
Additionally, an analysis of the degenerate problem of demixing with a single graph is also presented. 
Numerical experiments with synthetic and real-world graphs empirically illustrating the main theoretical findings close the paper.
\end{abstract}


\section{Introduction}\label{S:Introduction}

Modern interconnected systems and networks are growing more pervasive and complex by the day. While this is rendering data more abundant, its structure is more intricate than the one exhibited by classical datasets. Relying on graphs to model irregular information domains has emerged as a successful and tractable approach to deal with some of the challenges present in contemporary datasets. This approach, which has a long history in statistics \cite{kolaczyk2009book}, has been recently gaining traction in the machine learning \cite{hamilton2017representation,hu2020open} and signal processing communities \cite{djuric2018cooperative}. \textcolor{blue}{In the latter case, graph signal processing (GSP) puts forth a systematic approach where variables are modeled as nodes, the information domain is modeled as a graph (capturing heterogeneous relations from explicit pairwise relations), and the data itself is modeled as graph signals.} Graph signals are abundant in engineering and nature, including, for example, the congestion in traffic networks, the electrical neurological activity in a brain network, and the folding rate on protein residue networks.

The key in GSP is to postulate meaningful and tractable: i) models relating the properties of the data with the properties of the graph and ii) graph-signal operators that depend on the topology of the graph \cite{EmergingFieldGSP,ortega2018graph,marques2020graph}. With these two ingredients, a number of GSP-related problems have been investigated, ranging from basic sampling and reconstruction schemes \cite{SamplingOrtegaICASSP14,chen2015discrete,marques2016sampling,isufi2019forecasting,hara2022graph} to advanced deep graph neural network architectures \cite{bronstein2017geometric, defferrard2016convolutional, gama2019convolutional,gama2020graphs}. An area of particular interest is that of linear graph filters and their application to model linear network dynamics and solve inverse GSP-related problems. Mathematically, graph filters are graph-signal operators that can be represented by a matrix that is a polynomial of either the adjacency or Laplacian matrix of the graph \cite{djuric2018cooperative,EmergingFieldGSP,SandryMouraSPG_TSP13}. Practically, they can be implemented distributedly, account for a multi-resolution (multi-hop) description of the graph, and can capture spatio-temporal dynamics taking place across the nodes (agents) of a network \cite{SandryMouraICASSP14consensus,segarra2017optimal,coutino2019advances}. As a result, there has been a growing interest in using graph filters to postulate generative models of graph signals \cite{djuric2018cooperative,marques2017stationary,marques2020signal}. An approach is to model the signal at hand as the output of a graph filter excited with a parsimonious (e.g., sparse, white, or bandlimited) input graph signal. One can then formulate a number of inverse problems where (an imperfect version of) the output graph signal is given and, depending on the particular setup and the variables that are known, the goal is to recover the filter, the input, the parts of the output that are unknown or any combination thereof \cite{marques2020signal}. When the input graph signal is sparse (i.e., there are only a few nodes -sources- with a non-zero value), the associated output is referred to as a diffused graph signal, with sampling, reconstruction, blind deconvolution and filter identification of diffused graph signals having been investigated in detail \cite{segarra2016blind,zhu2020estimating,ramirez2017graph,thanou2017learning}.

\medskip\noindent\textbf{Contributions.} In this paper, we focus on blind demixing of graph signals. 
In particular, the observed graph signal $\by$ is assumed to be the combination (sum) of the diffused graph signals $\{\by_\sourceIndex\}_{\sourceIndex=1}^\numSources$, each of which is modeled as the output of a graph filter with coefficients $\bh_\sourceIndex$ excited with an input signal $\bx_\sourceIndex$.
The signal $\bx_\sourceIndex$ is assumed to be \emph{sparse}, localized in a subset of the nodes of the $\sourceIndex$th graph. \textcolor{blue}{In this setting, we seek to demix the observed signal $\by=\sum_\sourceIndex\by_\sourceIndex$ by \emph{jointly} estimating the $\{\bx_\sourceIndex\}_{\sourceIndex=1}^\numSources$ \emph{as well as} the $\{\bh_\sourceIndex\}_{\sourceIndex=1}^\numSources$.} This extends classical blind source separation (BSS) or demixing of signals in the time and spatial domains. Apart from its theoretical interest, identification of diffused graph signals also bears practical relevance, since they have been shown to be germane to, e.g., explaining the structure of real-world datasets \cite{rey2019sampling,zhu2020estimating} as well as understanding the dynamics of linear network diffusion processes \cite{segarra2016blind,zhu2020estimating,djuric2018cooperative}. 

\medskip\noindent\textbf{Target applications.} \textcolor{blue}{Although our contributions are biased toward modeling, algorithms and theory, their potential application is certainly important. Signal separation in networks bears practical relevance ranging from opinion formation and spread (e.g., in politics), to mean-field models for epidemics, to tracking the dynamics of physical phenomena (e.g., deforestation) across geographical networks, among others \cite{li2022learning,einizade2020detecting}.}

\medskip\noindent\textbf{Related work.} We generalize the problem of blind demixing to \emph{graph} filters and signals, providing \emph{both} algorithms and theoretical analysis. \textcolor{blue}{Relative to \cite{segarra2016blind}, which deals with the blind deconvolution of a \emph{single} graph signal (diffused by a single graph filter), demixing deals with multiple graph signals (each diffused by its graph filter) with the goal of separating them from a mixture.} Demixing is intrinsically more challenging than identification since it is a more undetermined problem. Relative to the classical (non-graph) demixing setup in \cite{ling2017blind}, the approach in our paper not only can be used in applications dealing with networks and irregular information domains but also entails different technical modeling assumptions \textcolor{blue}{(cf. last paragraph in Section~\ref{S:Recovery})}. \textcolor{blue}{Recent publications about demixing on networks include a deep learning approach for non-linear processes~\cite{vcutura2021deep}, extensions of classical ``JADE'' and ``ICA'' BSS methods to GSP~\cite{miettinen2021graph}, and methodologies based on signal smoothness and spectrally-constrained signals~\cite{mohammadi2023graph}.}

\medskip\noindent\textbf{Paper outline.} \Cref{S:fundamentals} summarizes general GSP concepts related to graph filters and linear diffusion network processes. \Cref{S:BlindDem} formulates the problem of blind demixing graph signals and proposes an efficient convex relaxation. Section \ref{S:Recovery} provides analytical results on the demixing performance, along with graph-specific parameters that affect the recovery guarantees. \Cref{S:demixing_single} deals with the special case of demixing with a single graph \emph{common} to all the mixing components. Numerical experiments illustrating the merits of our approach and the practical relevance of the theoretical analysis are presented in \Cref{S:Simulations}.\footnote{\noindent\textbf{Notational conventions:} The entries of a matrix $\bX$ and a (column) vector $\bx$ are denoted as $X_{ij}$ and $x_i$ or $[\bx]_i$. Operators $(\cdot)\transpose$, $(\cdot)\hermitian$, $\evn{}$, $\circ$, $\otimes$ and $\odot$ stand for matrix transpose, conjugate transpose (Hermitian), expectation, Hadamard (entry-wise), Kronecker, and Khatri-Rao (column-wise Kronecker) products, respectively. The complex conjugate of $x$ is denoted by $\bar{x}$; $\diag(\mathbf{x})$ is a diagonal matrix whose $(i,i)$th entry is equal to $x_i$; {$\mathcal{C}$ denotes a generic numerical constant}; and $|\cdot|$ is used for the cardinality of a set, and the magnitude of a scalar. The $N\times N$ identity matrix is represented by $\bI_N$, while $\mathbf{0}_N$ stands for the $N\times 1$ vector of all zeros, and $\mathbf{0}_{N\times P}=\mathbf{0}_N\mathbf{0}_P\transpose$. The sum of the absolute values of the elements of a vector $\bx$ is denoted as the norm $\| \bx \|_1$. In the case of matrices, the $\| \bX \|_{2,1}$ norm is the sum of the Euclidean norms (i.e. $\|\cdot\|_2$) of the rows of $\bX$. The number of non-zero elements of a vector $\bx$ is denoted as $\|\bx \|_0$. $\|\bX \|_\ast$ denotes the nuclear norm of matrix $\bX$ and $\|\bX \|_{2,0}$ its number of non-zero rows. \textcolor{blue}{The infinity and $\ell_2$ matrix norms, $\| \bX \|_\infty$ and $\| \bX \|$, stand for the entrywise largest absolute value and the largest singular value of $\bX$, respectively.} For a linear operator $\mathcal{X}$, $\| \mathcal{X} \| := \sup_{\bZ \neq 0} \| \mathcal{X}(\bZ) \|_\mathrm{F} / \| \bZ \|_\mathrm{F}= \| \bX \|$, where $\bX$ is the matrix representation of $\mathcal{X}$. Otherwise, standard vector and matrix norm notation is used.}

\section{Graph Filtering Fundamentals}\label{S:fundamentals}

Let $\mathcal{G}$ denote a directed graph with a set of nodes $\mathcal{N}$ (with cardinality $\numNodes$) and a set of links $\mathcal{E}$ such that if $i$ is connected to $j$ then $(i,j)\in\mathcal{E}$. Local connectivity is captured by the set $\mathcal{N}_i:=\{j\;|(j,i)\in\mathcal{E}\}$, which stands for the (incoming) neighborhood of $i$. For any given $\mathcal{G}$, we define the adjacency matrix $\bA\in\mathbb{R}^{\numNodes\times \numNodes}$ as a sparse matrix with non-zero elements $A_{ji}$ if and only if $(i,j)\in\mathcal{E}$. The value of $A_{ji}$ captures the strength of the connection from $i$ to $j$. 
The focus of the paper is on analyzing and modeling (graph) signals defined on $\mathcal{N}$. These signals can be represented as vectors $\mathbf{x}=[x_1,...,x_\numNodes]\transpose \in  \mathbb{R}^\numNodes$, where $x_i$ represents the value of the signal at node $i$.

\subsection{Graph-shift operator}\label{s:gsp_gso} Since the vector representation $\mathbf{x}=[x_1,...,x_\numNodes]\transpose$ does not account explicitly for the network structure, GSP endows $\mathcal{G}$ with the \emph{graph-shift operator} (GSO) $\mathbf{S}$ \cite{SandryMouraSPG_TSP13,djuric2018cooperative}. The shift $\mathbf{S}\in\mathbb{R}^{\numNodes \times \numNodes}$ is a matrix whose entry $S_{ji}$ can be non-zero only if $i=j$ or if $(i,j)\in\mathcal{E}$. The sparsity pattern of the matrix $\bS$ captures the local structure of $\mathcal{G}$, with no specific assumptions on the values of its non-zero entries.
The intuition behind $\mathbf{S}$ is to represent a linear transformation that can be computed locally at the nodes of the graph. More rigorously, if $\mathbf{y}$ is defined as $\mathbf{y}=\mathbf{S}\mathbf{x}$, then node $i$ can compute $y_i$ as linear combination of the signal values $x_j$ where $j\in \mathcal{N}_i$.
Typical choices for $\mathbf{S}$ are the adjacency matrix $\bA$ \cite{SandryMouraSPG_TSP13} and the graph Laplacian \cite{EmergingFieldGSP}. We assume henceforth that $\bS$ is diagonalizable, so that $\bS=\bV\bLambda\bV^{-1}$ with $\bLambda\in\mathbb{C}^{\numNodes \times \numNodes}$ being diagonal. 
In particular, $\bS$ is guaranteed to be diagonalizable when it is normal, i.e., when it satisfies $\bS\bS\hermitian=\bS\hermitian\bS$ (this is one of the core assumptions in \Cref{S:Recovery}). 
In that case, we have that $\bV$ is unitary, which implies $\bV^{-1}=\bV\hermitian$, and leads to the decomposition $\bS=\bV\bLambda\bV\hermitian$.

\subsection{Graph filters and diffusion processes} The shift $\bS$ can be used to define \emph{linear graph-signal operators} of the form
\begin{equation}\label{E:Filter_input_output_time}
\mathbf{H}:=\sum_{\tapIndex=0}^{\numTaps-1}h_\tapIndex \bS^\tapIndex,
\end{equation}
which are called \emph{graph filters}~\cite{SandryMouraSPG_TSP13,segarra2017optimal}. For a given input $\bx$, the output of the filter is simply $\by=\bH\bx$. The coefficients of the filter are collected into $\mathbf{h}:=[h_0,\ldots,h_{\numTaps-1}]\transpose$, with $\numTaps-1$ denoting the filter degree. Graph filters are of particular interest because they model network diffusion processes and can be implemented in a distributed fashion~\cite{segarra2017optimal,coutino2019advances}. 
Note also that the Cayley-Hamilton theorem guarantees that even infinite-horizon processes (i.e., those depending on arbitrarily large powers of $\bS$) can be rewritten as a filter of at most degree $\numNodes-1$.

\subsection{Spectral representation}\label{s:gsp_frequency} 
Leveraging the spectral decomposition of $\bS$, graph filters and signals can be represented in the frequency domain.
To be precise, let us use the eigenvectors of $\bS$ to define the $\numNodes\times \numNodes$ matrix $\bU:=\bV^{-1}$, and the eigenvalues of $\bS$ to define the $\numNodes \times \numTaps$ Vandermonde matrix  $\bPsi$, where $\Psi_{ij}:=(\Lambda_{ii})^{j-1}$.
Using these conventions, the frequency representations of a \emph{signal} $\bx$ and of a \emph{filter} $\bh$ are defined as $\widehat{\bx}:=\bU\bx$ and $\widehat{\bh}:=\bPsi\bh$, respectively~\cite{djuric2018cooperative,segarra2017optimal}. 
Exploiting such representations, the output $\by\!=\!\bH\bx$ of a graph filter in the frequency domain is given by
\begin{equation}\label{E:Filter_input_output_freq}
	\widehat{\by}=\diag\big(\bPsi\bh\big)\bU \bx=\diag\big(\widehat{\bh}\big)\widehat{\bx}=
	\widehat{\bh}\circ\widehat{\bx},
\end{equation}
with $\circ$ being the Hadamart product. Identity \eqref{E:Filter_input_output_freq} is the counterpart of the celebrated convolution theorem for temporal signals, and follows from $\mathbf{H}=\bV\big(\sum_{\tapIndex=0}^{\numTaps-1}h_\tapIndex \boldsymbol{\Lambda}^\tapIndex\big)\bU$ [cf. \eqref{E:Filter_input_output_time}] and $\sum_{\tapIndex=0}^{\numTaps-1}h_\tapIndex \boldsymbol{\Lambda}^l=\diag{(\bPsi\bh)}$. Having  $\widehat{\by}=\widehat{\bh}\circ\widehat{\bx}$ implies that the dependence of $\widehat{\by}$ on the frequency representations of $\bh$ and $\bx$ is bilinear, with no mixing among frequencies (i.e, $[\widehat{\by}]_k$ depends only on $[\widehat{\bh}]_k[\widehat{\bx}]_k$). In contrast, if one wants to write $\widehat{\by}$ as a function of the entries of $\bh$ and $\bx$, it follows from \eqref{E:Filter_input_output_freq} that the expression is
\begin{equation}\label{E:Filter_input_output_Ofreq_FInode}
\widehat{\by}=(\bPsi_\sourceIndex\transpose \odot \bU_\sourceIndex\transpose )\transpose\vect(\bx\bh\transpose),
\end{equation}
with $\odot$ being the Khatri-Rao product; see, e.g, \cite{segarra2016blind}. That is, the dependency is still bilinear, but any entry of $\widehat{\by}$ depends on all the bilinear terms $[\bh]_\tapIndex [\bx]_n$. Interestingly, while in the time domain $\boldsymbol{\Psi}=\bU$, in GSP the matrices representing the graph Fourier transform (GFT) for filters and signals are in general not the same.

\section{Blind Demixing with Multiple Diffusing Graphs}\label{S:BlindDem}

Given an observation $\by=\sum_{\sourceIndex=1}^\numSources\by_\sourceIndex$, we aim to identify the $\numSources$ individual signals under the assumption that each $\by_\sourceIndex$ can be modeled as a diffused graph signal. That is, $\by_\sourceIndex$ is the output obtained by processing a \emph{sparse} input signal $\bx_\sourceIndex$ of \emph{unknown} support $\mS_\sourceIndex$ (of cardinality $|\mS_\sourceIndex| = \numNonzero_\sourceIndex$) with a graph filter of unknown coefficients $\bh_\sourceIndex$ and known GSO $\bS_\sourceIndex$. Formally, by modeling each $\by_\sourceIndex$ as
\begin{equation}\label{E:diffused_graph_signal_generative_model}
\by_\sourceIndex=\sum_{\tapIndex=0}^{\numTaps-1}[\bh_\sourceIndex]_\tapIndex \bS_\sourceIndex^\tapIndex \bx_{\sourceIndex},\;\;\text{where}\;\; [\bx_\sourceIndex]_n=0\;\;\nodeIndex\notin \mS_\sourceIndex,
\end{equation}
we look at the problem of recovering $\{\bx_{\sourceIndex}, \bh_{\sourceIndex}, \by_{\sourceIndex} \}_{\sourceIndex=1}^\numSources$ from the observed $\by$. 

With $\bPsi_\sourceIndex$, $\bV_\sourceIndex$ and $\bU_\sourceIndex$ denoting the (inverse) GFT matrices associated with the shift $\bS_\sourceIndex$ (cf. \Cref{s:gsp_gso}), and leveraging the spectral representation of graph signals and filters (cf. \Cref{s:gsp_frequency}), the joint demixing and blind deconvolution problem can be written as 
\begin{subequations}\label{E:intuitive_demixing_formulation_nolifting}
	\begin{align}
	\mathop{\text{find}} &\quad  \{\bx_{\sourceIndex}, \bh_{\sourceIndex}, \by_{\sourceIndex} \}_{\sourceIndex=1}^\numSources \\
	\label{E:intuitive_demixing_formulation_nolifting.line1} \text{s. to} &\quad  \by_\sourceIndex=\bV_\sourceIndex \left(\bPsi_\sourceIndex\transpose \odot \bU_\sourceIndex\transpose \right)\transpose\vect(\bx_\sourceIndex \bh_\sourceIndex\transpose )\;\;\;\forall \sourceIndex \\
	\label{E:intuitive_demixing_formulation_nolifting.line2} &\quad \by = {  \textstyle \sum_{\sourceIndex=1}^\numSources} \by_\sourceIndex, \quad \quad \|\bx_\sourceIndex \|_0\leq \numNonzero_\sourceIndex \;\;\;\forall \sourceIndex,
	\end{align}
\end{subequations}
\textcolor{blue}{where s. to stands for subject to.} When solving the above optimization, we assume that we know: i) the GSOs $\{\bS_\sourceIndex\}_{\sourceIndex=1}^\numSources$, so that $\{\bPsi_\sourceIndex, \bV_\sourceIndex, \bU_\sourceIndex\}_{\sourceIndex=1}^\numSources$ are available; and ii) the full vector of observations $\by$ (see \Cref{R:robust_demixing} for partially observable setups).

To handle the demixing problem in \eqref{E:intuitive_demixing_formulation_nolifting}, note first that the constraint \eqref{E:intuitive_demixing_formulation_nolifting.line1} implies that signals $\by_\sourceIndex$ are fully determined by $\bx_\sourceIndex$ and $\bh_\sourceIndex$. Hence, \eqref{E:intuitive_demixing_formulation_nolifting.line1} can be combined with the pertinent constraint in \eqref{E:intuitive_demixing_formulation_nolifting.line2}, so that $\{\by_\sourceIndex\}_{\sourceIndex=1}^\numSources$ are no longer explicit optimization variables, but obtained using \eqref{E:diffused_graph_signal_generative_model} after the optimization over $\bx_{\sourceIndex}$ and $\bh_{\sourceIndex}$ has been solved. Although this reduces considerably the dimension of the feasible set, the optimization is still challenging for a number of reasons. Firstly, even if the support of the inputs were known, for the problem to be well-posed, the number of non-trivial unknowns $\sum_{\sourceIndex=1}^\numSources(\numNonzero_\sourceIndex + \numTaps_\sourceIndex)$ needs to be less than the number of observations $\numNodes$. Secondly, each of the terms $\bx_\sourceIndex \bh_\sourceIndex\transpose$ is bilinear, which introduces a source of non-convexity and gives rise to an inherent scaling ambiguity. Thirdly, the presence of the $\ell_0$ norm renders the optimization NP-hard.
	
The first step to design a tractable relaxation of \eqref{E:intuitive_demixing_formulation_nolifting}, is to \textit{lift} the problem by defining the $\numNodes \!\times\! \numTaps_\sourceIndex$ rank-one matrices $\bZ_\sourceIndex\!=\!\bx_\sourceIndex\bh_\sourceIndex\transpose$, together with the $\numNodes \!\times\! \numNodes \numTaps_\sourceIndex$ transfer matrices $\bM_\sourceIndex=\bV_\sourceIndex \left(\bPsi_\sourceIndex\transpose \odot \bU_\sourceIndex\transpose \right)\transpose$. With these notational conventions, and combining the \textit{linear} constraints in \eqref{E:intuitive_demixing_formulation_nolifting.line1}-\eqref{E:intuitive_demixing_formulation_nolifting.line2} into a single one, the demixing problem can be equivalently written as 
\begin{align}
\label{eq:feasibility2}
\{\bZ_\sourceIndex^{(*)}\}_{\sourceIndex=1}^\numSources & =\text{find} \quad \{\bZ_\sourceIndex\}_{\sourceIndex=1}^\numSources \\
\text{s. to} \,\,\,\, \by & \!=\! {  \textstyle \sum_{\sourceIndex=1}^\numSources} \bM_\sourceIndex \text{vec}\!\left(\bZ_\sourceIndex\right),\;\,  \text{rank}(\bZ_\sourceIndex) \!= \!1, \;\, \|\bZ_\sourceIndex\|_{2,0} \leq \numNonzero_\sourceIndex. \nonumber
\end{align}
Note that $\|\bZ_\sourceIndex\|_{2,0}$, defined as the number of non-zero rows of $\bZ_\sourceIndex$, mimics the role of $\|\bx_\sourceIndex\|_0$ in \eqref{E:intuitive_demixing_formulation_nolifting}. The solution vectors $\bx_\sourceIndex^{(*)}$ and $\bh_\sourceIndex^{(*)}$ are given by the scaled versions of the right and left principal \textit{singular vectors} of the rank-one matrix $\bZ_\sourceIndex^{(*)}$. Although the equivalent formulation \eqref{eq:feasibility2} is still difficult to solve, it leads to a convex relaxation which is described in the ensuing section, after the following remark.
	
\begin{remark}[Robust demixing] \normalfont \label{R:robust_demixing}
	\label{remark:robust_demixing}
	The demixing problem -- in particular constraint \eqref{E:intuitive_demixing_formulation_nolifting.line1} -- can be rewritten to account for noisy and incomplete observations. If the values of $\by$ are observed only at a subset of $\numSamples$ nodes, both sides of \eqref{E:intuitive_demixing_formulation_nolifting.line1} must be left multiplied by the \textit{sampling matrix} $\bC_{\mathcal{M}}\! \in \! \{0,1\}^{\numSamples \times \numNodes}$ whose rows correspond to canonical basis vectors identifying the indexes $\mathcal{M}=\{n_m\}_{m=1}^\numSamples$ of the observed entries. Presence of noise in the observations would require replacing the equality in \eqref{E:intuitive_demixing_formulation_nolifting.line1} with a term penalizing the difference between the two sides of \eqref{E:intuitive_demixing_formulation_nolifting.line1}, with the penalty depending on the specific type of noise.
\end{remark}
	
\subsection{Convex relaxation and algorithmic approach}\label{SS:Convex_relax_multiple}
The \emph{feasibility} problem in \eqref{eq:feasibility2} is reformulated as a rank-\textit{minimization} problem where, first, the objective is augmented with $\ell_{2,0}$ constraints and, then, tractable surrogates of the (non-convex) rank and the $\ell_{2,0}$ norm are used. Specifically, consider replacing the rank with the nuclear norm, and the $\ell_{2,0}$ norm with the $\ell_{2,1}$ norm~\cite{Fazel2001,Tropp2006}. The resultant optimization is
\begin{align}
\label{eq:convex_minization2}
\{\bZ_\sourceIndex^{(*)}\}_{\sourceIndex=1}^\numSources & = \argmin_{\{\bZ_\sourceIndex\}_{\sourceIndex=1}^\numSources} \quad  \textstyle \sum_{\sourceIndex=1}^\numSources \eta_{\sourceIndex}\|\bZ_\sourceIndex\|_* + \sum_{\sourceIndex=1}^\numSources \beta_{\sourceIndex}\|\bZ_\sourceIndex\|_{2,1} \\
& \text{s. to} \quad  \by = {  \textstyle \sum_{\sourceIndex=1}^\numSources \bM_\sourceIndex} \text{vec}\left(\bZ_\sourceIndex\right),\nonumber  
\end{align}
where  $\{\eta_{\sourceIndex},\beta_{\sourceIndex}\}_{\sourceIndex=1}^\numSources$ are tuning constants that, if no prior information is available, can be set to $\eta_{\sourceIndex} = 1$ and $\beta_{\sourceIndex}=\beta$ for all $\sourceIndex$. The optimization in \eqref{eq:convex_minization2} is convex, so that it can be handled with generic off-the-shelf solvers. When computational complexity is a concern, one can alternatively develop efficient algorithms tailored to the structure of \eqref{eq:convex_minization2}, see, e.g., \cite{recht2010guaranteed} for a related problem.

Recoverability conditions can be found by ignoring the low-rank nature of all the $\bZ_\sourceIndex$ and focusing on the fact that the concatenation of the vectorized forms of all these matrices $\bz = \mathrm{vec}([\bZ_1, \ldots, \bZ_\numSources])$ is effectively a sparse vector. In this way, one can think of recovering a sparse vector $\bz$ satisfying $\by = [\bM_1, \ldots, \bM_\numSources] \, \bz$ and derive recoverability conditions in this basic setting~\cite{book_comp_sensing}. A more sophisticated analysis accounting for the structure of the linear operators $\bM_\sourceIndex$ and the low-rank nature of $\bZ_\sourceIndex$ is our main theoretical contribution, which is presented in \Cref{S:Recovery}.

\section{Exact Blind Demixing Via Nuclear Norm Minimization}\label{S:Recovery}

This section presents our main theoretical result, namely, under a few technical conditions (to be outlined next) all the filters
$\bh_\sourceIndex$ and sparse input signals $\bx_\sourceIndex$ in \eqref{E:intuitive_demixing_formulation_nolifting} can be exactly recovered 
with high probability by solving a simplified version of the convex problem \eqref{eq:convex_minization2} where we only keep the nuclear norm in the objective, i.e.,
\begin{equation}
\label{eq:nuclear_optimization}
\textcolor{blue}{\text{min}} \sum_{\sourceIndex=1}^\numSources \|\bZ_\sourceIndex\|_* \quad \textcolor{blue}{\text{s. to}} \quad \by=\sum_{\sourceIndex=1}^\numSources \bM_\sourceIndex \text{vec}\left(\bZ_\sourceIndex\right).
\end{equation}
\textcolor{blue}{The objective under minimization is simplified from \eqref{eq:convex_minization2} to \eqref{eq:nuclear_optimization} to enable the analysis in the Appendix, which results in Theorem~\ref{T:main_theorem}. Nonetheless, while not promoted explicitly in \eqref{eq:nuclear_optimization}, the colum sparsity of $\bZ_\sourceIndex$ will play a role in the recovery guarantees (see the Appendix for details).}

\textcolor{blue}{To prove the recovery result stated formally in Theorem~\ref{T:main_theorem}, we employ \emph{concentration functions} to operate on the GFT matrices for filters and signals} (cf. Section \ref{s:gsp_frequency}). In~\cite{segarra2016blind}, given an arbitrary matrix $\bA \in \mathbb{C}^{M \times N}$ and a positive integer $k \leq  N$, the function $\rho$ is defined as
\begin{equation}
	\label{eq:rho}
	\rho_\bA^{(k)} := \max_{1 \le \ell \le M} \max_{\Omega \in \Omega_k^N} \|\ba_{\ell_\Omega}\|_2^2,
\end{equation}
where $\Omega_k^N$ represents the set of all $k$-subsets of $\{1,2,\dots,N\}$ and $\ba_{\ell_\Omega}$ the orthogonal projection of the $\ell$th row of $\bA$ onto the index set $\Omega$.
Building on this idea, let us now consider a second matrix $\bB \in \mathbb{C}^{M \times N}$ and, with positive integers $k_1$ and $k_2$, define the function $\kappa$ as
\begin{equation}
	\label{eq:kappa}
	\kappa_{\bA, \bB}^{(k_1, k_2)} := \max_{1 \le \ell \le M} \max_{(\Omega_1, \Omega_2) \in \Omega_{k_1}^N \times \Omega_{k_2}^N} \|\ba_{{\ell}_{\Omega_1}} \ \bb_{{\ell}_{\Omega_2}}\hermitian\|.
\end{equation}
In order to relate $\rho$ and $\kappa$ recall that the matrix norm $\|\bA\|$ is induced from the vector norm $\sup_{\bx}\|\bA \bx\|_2$ where $ \|\bx\|_2=1$. If the matrix under consideration is of rank one, as in \eqref{eq:kappa}, it can be written as $\bA = \ba \bb\hermitian$, whose norm is given in terms of $\|\ba \bb\hermitian \bx\|_2$. Organizing the terms, $\|\ba \bb\hermitian \bx\|_2 = |\bb\hermitian \bx| \ \|\ba\|_2$, which is maximum when $\bx$ is parallel to $\bb$ and, therefore, $\|\ba \bb\hermitian\| = \|\ba\|_2 \|\bb\|_2$. Thus, we can relate $\rho$ with $\kappa$ as
\begin{equation}
	\label{eq:kappa_rho_relation}
	\kappa_{\bA, \bB}^{(k_1, k_2)} \le \sqrt{\rho_\bA^{(k_1)}} \sqrt{\rho_\bB^{(k_2)}}.
\end{equation}

\textcolor{blue}{The assumptions put in place in our analysis are outlined in Section \ref{Ss:Scope}, after which the main result is formally stated in Section~\ref{Ss:Results} and proved in the Appendix}.

\subsection{Assumptions and scope of the analysis}\label{Ss:Scope}

We start by introducing two assumptions that render the analysis more tractable.

\noindent \textbf{Assumption 1 (AS1)}  The GSOs $\bS_\sourceIndex, \ \sourceIndex\!\in\!\{1,2,\dots,\numSources\}$, are normal, i.e., they satisfy $\bS_\sourceIndex\bS_\sourceIndex\hermitian=\bS_\sourceIndex\hermitian\bS_\sourceIndex$, and for each $\bS_\sourceIndex$ its eigenvalues are distinct.

\noindent \textbf{Assumption 2 (AS2)}  \textcolor{blue}{The sparse signals' supports $\mS_\sourceIndex, \sourceIndex\!\in\!\{1,2,\dots,\numSources\}$ are known. In other words, with $\bC_{\mS_\sourceIndex}\transpose$ being a tall matrix whose rows are either all zeros, or rows from the identity matrix, the mapping $\bx_\sourceIndex = \bC_{\mS_\sourceIndex}\transpose \bs_\sourceIndex$ between the sparse signals $\bx_\sourceIndex$ and the ``seeding'' signals $\bs_\sourceIndex$ is known. }

\noindent \textbf{Assumption 3 (AS3)} The observation is a mix of filtered graph signals and its frequency representation adheres to the model
\begin{equation}
\label{eq:proof_observation_model}
\textcolor{blue}{\widehat{\bnu}=\sum_{\sourceIndex=1}^\numSources\diag(\bPsi_\sourceIndex \bh_\sourceIndex) \widetilde{\bU}_\sourceIndex\bx_\sourceIndex=\sum_{\sourceIndex=1}^\numSources\diag(\bPsi_\sourceIndex \bh_\sourceIndex) \bA_\sourceIndex \bs_\sourceIndex,}
\end{equation}
\textcolor{blue}{where $\bA_\sourceIndex := \widetilde{\bU}_\sourceIndex \bC_{\mS_\sourceIndex}\transpose \in \mathbb{C}^{\numNodes \times \numNonzero_\sourceIndex}$ and $\widetilde{\bU}_\sourceIndex$ is a random $\numNodes \times \numNodes$ matrix obtained by concatenating $\numNodes$~rows sampled independently and uniformly with replacement from~$\bU_\sourceIndex$.}

Under (AS1) the matrices $\bV_\sourceIndex$ are unitary, which implies $\bU_\sourceIndex:=\bV_\sourceIndex^{-1}=\bV_\sourceIndex\hermitian$ and leads to the decomposition $\bS_\sourceIndex=\bV_\sourceIndex\bLambda_\sourceIndex\bV_\sourceIndex\hermitian$. Normality is for instance satisfied when $\mathcal{G}$ is undirected and the GSO is chosen to be the adjacency matrix or the graph Laplacian. Furthermore, that each $\bS_\sourceIndex$ has distinct eigenvalues ensures that the Vandermonde matrices $\bPsi_\sourceIndex$ are full-rank independently of $\numTaps_\sourceIndex$, which is required for uniqueness.

\textcolor{blue}{Assuming the supports $\mS_\sourceIndex$ are known via (AS2) enables reducing the degrees of freedom of $\bx_\sourceIndex$. Signals $\bx_\sourceIndex$ belonging to a vector subspace of dimension $\numNonzero_\sourceIndex$ smaller than $\numNodes$ will allow us to obtain appropriate bounds.}

Concerning the probabilistic model for the observations in (AS3), this class of models is customary when establishing recovery guarantees of sparse signal reconstruction and low-rank matrix completion in, e.g., compressive sensing~\cite{book_comp_sensing} and demixing of classical signals~\cite{ling2017blind,flinth2018sparse}. \textcolor{blue}{In particular, the application of Bernstein inequalities for random matrices is instrumental to the proof of Theorem~\ref{T:main_theorem} (see the Appendix).}

\subsection{Main theoretical result}\label{Ss:Results}

Consider $\numSources$ $\numNonzero_\sourceIndex$-sparse graph signals $\bx_\sourceIndex\!\in\!\mathbb{R}^\numNodes, \sourceIndex\!\in\!\{1,2,\dots,\numSources\}$, each respectively diffused through its supporting graph using a filter with coefficients $\bh_\sourceIndex \in\mathbb{R}^{\numTaps_\sourceIndex}$, and giving rise to an aggregated signal whose frequency representation adheres to the model in (AS3).
With the matrices $\bU_\sourceIndex\in\mathbb{C}^{\numNodes \times \numNodes}$ and $\bPsi_\sourceIndex\in\mathbb{C}^{\numNodes\times \numTaps_\sourceIndex}$ denoting the GFT for signals and filters associated with the GSOs $\bS_\sourceIndex$, respectively, we assume without loss of generality that the matrices $\bU_\sourceIndex$ are normalized so that $\bU_\sourceIndex\hermitian \bU_\sourceIndex = \numNodes \bI_\numNodes$. 
In this setting, the following result holds.
\begin{mytheorem}\label{T:main_theorem}
Under Assumptions (AS1)-(AS3), the unique solution to the convex program \eqref{eq:nuclear_optimization} coincides with the original search \eqref{eq:feasibility2} with probability at least $1-\numNodes^{-\alpha+1}$ where $1 \le \alpha \le \min\{\alpha_1, \alpha_2, \alpha_3\}$, $\rho_{\max}^{(\yetOtherSourceIndex,\otherSourceIndex)} = \max\left\{\rho_{\bU_\yetOtherSourceIndex}^{(\numNonzero_\yetOtherSourceIndex)}, \rho_{\bU_\otherSourceIndex}^{(\numNonzero_\otherSourceIndex)}\right\}$, and
\begin{equation*}
\begin{split}
	&\alpha_1 \! = \!\! \min_{\sourceIndex} \frac{3}{128} \left(\rho_{\bPsi_\sourceIndex}^{(\numTaps_\sourceIndex)} \rho_{\bU_\sourceIndex}^{(\numNonzero_\sourceIndex)} \log\left(2 \numNodes \numTaps_\sourceIndex \numNonzero_\sourceIndex\right)\right)^{-1} \\
	&\alpha_2 \! = \! \min_{\otherSourceIndex\neq \yetOtherSourceIndex}\frac{9}{32} \frac{\Big(\rho_{\bPsi_\yetOtherSourceIndex}^{(\numTaps_\yetOtherSourceIndex)} \rho_{\bPsi_\otherSourceIndex}^{(\numTaps_\otherSourceIndex)} \rho_{\max}^{(\yetOtherSourceIndex,\otherSourceIndex)} + \frac{1}{\numSources} \kappa_{\bU_{\yetOtherSourceIndex}, \bU_{\otherSourceIndex}}^{\left(\numNonzero_\yetOtherSourceIndex,\numNonzero_\otherSourceIndex\right)} \kappa_{\bPsi_{\yetOtherSourceIndex} \bPsi_{\otherSourceIndex}}^{\left(\numTaps_\yetOtherSourceIndex,\numTaps_\otherSourceIndex\right)}\Big)^{-1}}{\numSources^2\log(2\numNodes)} \\
	&\alpha_3 \! = \! \textcolor{blue}{\frac{1}{\mathcal{C}} \min_{\sourceIndex} \frac{\numNodes}{ \numSources (\rho_{\max} \! + \! 1)^2 \! \max\{\mu_{\max}^2 \numTaps_\sourceIndex, \mu_h^2 \numNonzero_\sourceIndex\} \log^2\numNodes} - \! \log\numSources} \\
\end{split}
\end{equation*}
\end{mytheorem}
\noindent \emph{Proof}: See \textcolor{blue}{the Appendix}.

\noindent Note that the statement of the theorem leverages the definitions in \eqref{eq:rho}-\eqref{eq:kappa} and depends on the scalar quantities $\mu_h$ and $\mu_{\max}$, which can be interpreted as generalizations of \eqref{eq:rho}-\eqref{eq:kappa} capturing the interactions between $\bh_\sourceIndex$ and $\bPsi_\sourceIndex$, and are defined in detail in \eqref{eq:muh}-\eqref{eq:mumax}.

The main goal of the probabilistic recovery bounds in Theorem~\ref{T:main_theorem} is to establish that a convex relaxation \emph{can} succeed in blind demixing of graph signals and, in practice, extrapolate from the theoretical results a \emph{predictor} of demixing performance based on the graph topologies. 
Although admittedly loose, the bounds in the theorem motivate empirical predictors of successful demixing as illustrated in Section~\ref{S:Simulations}.

Furthermore, we want to emphasize two differences between Theorem~\ref{T:main_theorem} and the main result in~\cite{ling2017blind}. First, Theorem~\ref{T:main_theorem} provides recovery guarantees for blind demixing in {\textit{arbitrary}} graphs, whereas \cite{ling2017blind} only applies to Gaussian $\widetilde{\bU}_\sourceIndex$. Second, \cite{ling2017blind} employs an explicit subspace model that does not leverage sparsity in the mixing signals, and instead introduces auxiliary linear subspaces. \textcolor{blue}{In contrast, we leverage the matrices $\bC_{\mS_\sourceIndex}\transpose$ to relate the explicit subspace to the sparse inputs model.}

\section{Blind Demixing With A Single Diffusing Graph}\label{S:demixing_single}

So far, we have considered the case where each input $\bx_{\sourceIndex}$ is diffused across its own graph characterized by its GSO $\bS_\sourceIndex$. 
The setting where all the signals are diffused in the same graph and then combined leads to a more challenging blind demixing problem. 
Notice that if $\bS_\sourceIndex=\bS$ for all $\sourceIndex$, then the transfer matrix $\bM_\sourceIndex=\bM$ is equivalent for all the inputs. As a result, the observation equation $\by = \sum_{\sourceIndex=1}^\numSources \bM_\sourceIndex \text{vec}( \bZ_\sourceIndex)$ simplifies to $\by = \bM \text{vec} (\sum_{\sourceIndex=1}^\numSources \! \bZ_\sourceIndex)$, which introduces a new source of ambiguity as explained in detail next.

\subsection{Single graph demixing ambiguity}\label{SS:ambiguity_single}

First, consider the minimization form of the feasibility problem in~\eqref{eq:feasibility2}, i.e., a non-relaxed predecessor of~\eqref{eq:convex_minization2},
\begin{align}
	\label{eq:feasibility2_singlegraph}
	\{\bZ_\sourceIndex^{(*)}\}_{\sourceIndex=1}^\numSources & = \argmin_{\{\bZ_\sourceIndex\}_{\sourceIndex=1}^\numSources} \quad  \textstyle \sum_{\sourceIndex=1}^\numSources \eta_{\sourceIndex} \, \text{rank}(\bZ_\sourceIndex) + \sum_{\sourceIndex=1}^\numSources \beta_{\sourceIndex}\|\bZ_\sourceIndex\|_{2,0} \nonumber \\
	\text{s. to} \quad & \by = \bM \, \text{vec}\!\left({  \textstyle \sum_{\sourceIndex=1}^\numSources \! \bZ_\sourceIndex}\right).
\end{align}
To illustrate the ambiguity emerging in the single graph demixing problem, consider the case where $\numSources\!=\!2$ and, accordingly, $\bZ_1^{(*)}$ and $\bZ_2^{(*)}$ denote the true rank-one matrices. 
Next, use these matrices to define $\bZ_1'=\bZ_1^{(*)}+\bZ_2^{(*)}$ and $\bZ_2'=\bzero$. It readily follows that: 1) both $\{\bZ_1^{(*)},\bZ_2^{(*)}\}$ and $\{\bZ_1',\bZ_2'\}$ are in the feasible set of \eqref{eq:feasibility2_singlegraph}; and 2) employing the triangle inequality: $\text{rank}(\bZ_1')+\text{rank}(\bZ_2')=\text{rank}(\bZ_1')\leq\text{rank}(\bZ_1^{(*)})+\text{rank}(\bZ_2^{(*)})$. In words, the true $\{\bZ_1^{(*)},\bZ_2^{(*)}\}$ involve a (rank) cost larger than or equal to that of $\{\bZ_1',\bZ_2'\}$. Furthermore, if $\eta_1\neq \eta_2$ then a solution where either $\bZ_1'=\mathbf{0}$ or $\bZ_2' = \mathbf{0}$ always achieves a smaller rank cost. A similar argument holds for the $\ell_{2,0}$ norm that constitutes the objective in \eqref{eq:feasibility2_singlegraph} together with the rank just analyzed.

This demonstrates that the demixing problem with a single diffusing graph presents additional challenges. With respect to the multiple-graph problem, the individual $\bZ_\sourceIndex$ matrices are in general non-identifiable separately, albeit the combination $\bZ \!= \!\sum_{\sourceIndex=1}^\numSources \! \bZ_\sourceIndex$ can be recovered. Thus, we propose a two-step approach in which we first efficiently obtain the true $\bZ^{(*)} \!= \!\sum_{\sourceIndex=1}^\numSources \! \bZ^{(*)}_\sourceIndex$ and then outline conditions for which the constituent $\bZ^{(*)}_\sourceIndex$ matrices can be uniquely recovered.

\subsection{Convex relaxation and algorithmic approach}\label{SS:Convex_relax_single}

Following the strategy in \Cref{SS:Convex_relax_multiple} to design an efficient demixing algorithm, the combinatorial norms in \eqref{eq:feasibility2_singlegraph} are replaced with their convex surrogates $\|\cdot\|_*$ and $\|\cdot\|_{2,1}$. This results in an instance of \eqref{eq:convex_minization2} with all $\bM_\sourceIndex=\bM$ whose solutions $\{\bZ_\sourceIndex^{(*)}\}_{\sourceIndex=1}^\numSources$ also suffer from the linear ambiguity present in single graph demixing. In fact, given that the nuclear as well as the $\ell_{2,1}$ norms are absolutely homogeneous (whereas the rank and the $\ell_{2,0}$ penalties are not), the ambiguity is more severe in the convex formulation. To see this, note that any solution of the form $\bZ_\sourceIndex\!=\!\theta_\sourceIndex\bZ^{(*)}$ with $\theta_\sourceIndex\!\geq\! 0$ and $\sum_{\sourceIndex=1}^\numSources\theta_\sourceIndex\!=\!1$ is feasible and yields the same value of  $\sum_{\sourceIndex=1}^\numSources\|\bZ_\sourceIndex\|_*$. Hence, the postulated optimization will have multiple solutions with the same cost than that of the actual $\bZ_\sourceIndex^{(*)}$ we aim to identify.

Another option consists in reformulating \eqref{eq:convex_minization2} based on $\bZ=\sum_{\sourceIndex=1}^\numSources\bZ_\sourceIndex$ as follows
\begin{equation}
	\label{eq:convex_minization_sum}
	\bZ^{(*)}  = \argmin_{\bZ} \,\,  \textstyle \|\bZ\|_* + \beta \|\bZ\|_{2,1}
	\quad \,\, \text{s. to} \,\,\,  \by = \textstyle \bM  \text{vec}\left(\bZ\right).
\end{equation}
In terms of computational complexity, this reformulation presents a lower cost because the number of optimization variables are effectively reduced by a factor of $\numSources$. Moreover, the formulation in \eqref{eq:convex_minization_sum} is similar to that of blind deconvolution \cite{segarra2016blind} (that is, a single source without demixing) with the exception that the true $\bZ^{(*)}$ to be recovered, albeit being low-rank, is not necessarily rank-one. Still, the results in~\cite[Theorem~1]{segarra2016blind} can provide probabilistic guarantees for the recovery of the true $\bZ^{(*)}$ in terms of the matrices $\bPsi$ and $\bU$ associated with the common diffusing graph.

\begin{figure*}
	\centering
	
	\begin{subfigure}{.31\textwidth}
		\centering
		\resizebox{\textwidth}{!}{\includegraphics{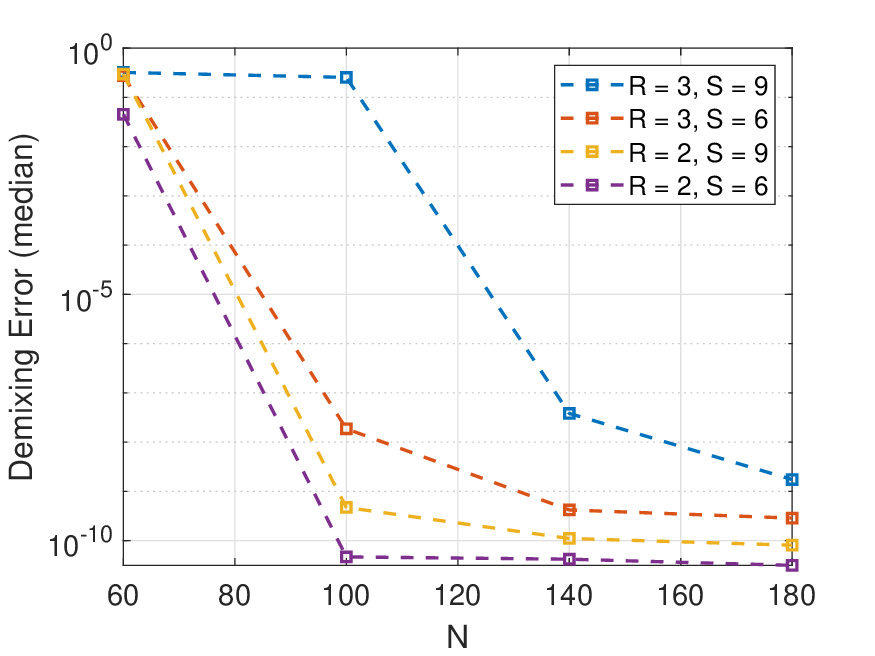}}
		\caption{}
		\label{fig:singlegraph}
	\end{subfigure}%
	\hspace{0.4cm}
	\begin{subfigure}{.31\textwidth}
		\centering
		\resizebox{\textwidth}{!}{\includegraphics{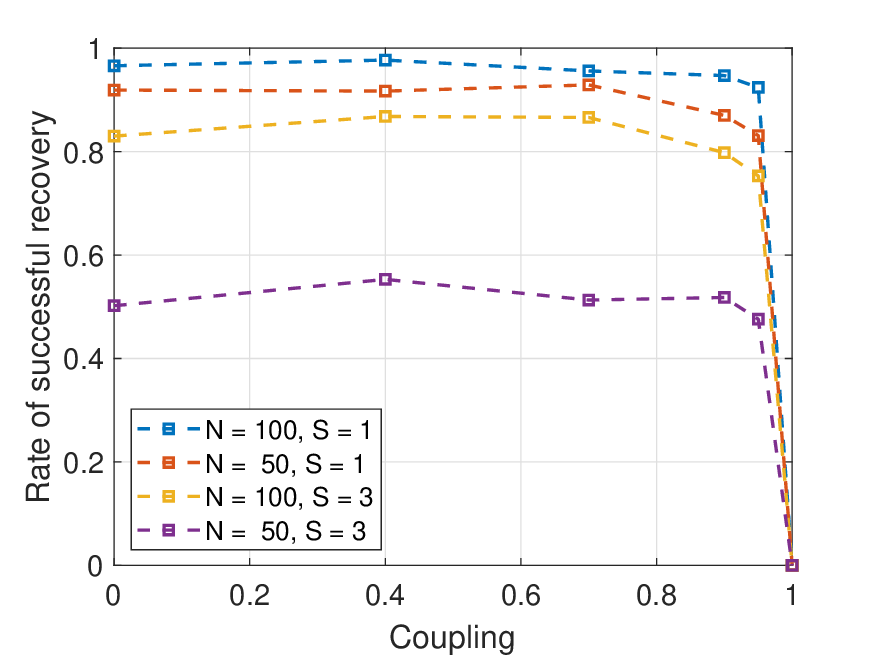}}
		\caption{}
		\label{fig:twograph}
	\end{subfigure}%
	\hspace{0.4cm}
	\begin{subfigure}{.31\textwidth}
		\centering
		\resizebox{\textwidth}{!}{\includegraphics{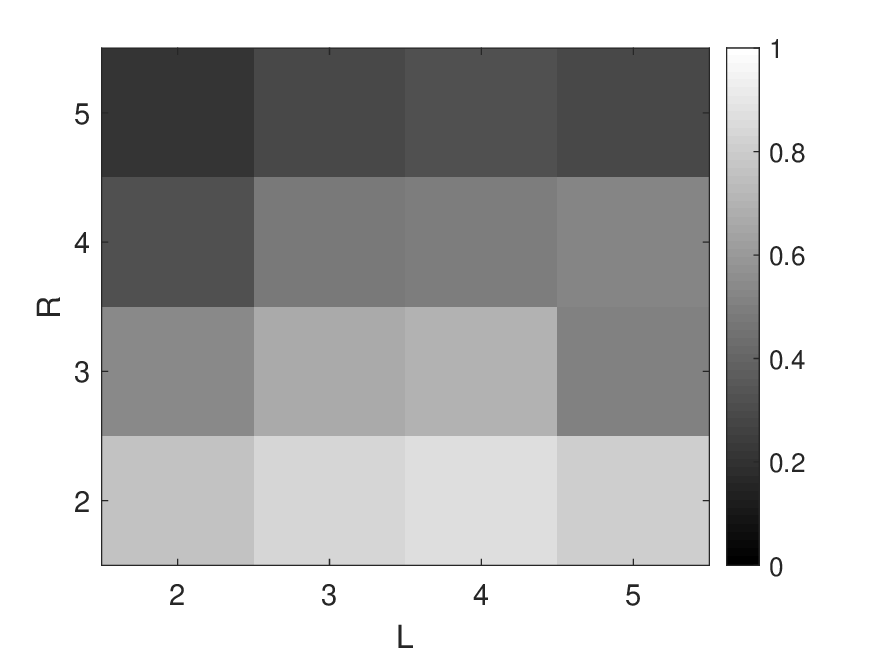}}
		\caption{}
		\label{fig:brain}
	\end{subfigure}
	
	\vspace{-0.1in}
	\caption{\small Blind demixing performance in single-graph and multiple-graph settings. (a)~Demixing error as a function of the graph size $\numNodes$ in a single graph for four settings of $\numNonzero$ and $\numSources$. (b)~Rate of successful recovery between a pair of graphs as a function of the \emph{similarity} parameter $\alpha$ between them (when $\alpha=0$ the two graphs are statistically independent, when $\alpha=1$ the graphs are the same). (c)~Rate of successful recovery for multiple brain graphs for several settings of $\numTaps$ and $\numSources$.}
	\vspace{-0.1in}
	\label{F:num_exp}
\end{figure*}

\subsection{Resolving the ambiguity}\label{SS:Ambiguity_single}

It is in general not feasible to recover $\{\bZ_\sourceIndex^{(*)}\}_{\sourceIndex=1}^\numSources$ from $\bZ^{(*)}$ due to the single graph ambiguity already explained. 
Consequently, we propose one way to ensure recoverability by means of imposing orthogonality conditions on the filter coefficients and the input signals. 
Specifically, we consider the following two separate assumptions:
\begin{description}
	\item[(a1)] $\bx_{\sourceIndex}\transpose \bx_{\otherSourceIndex}=0$ or $\widehat{\bx}_{\sourceIndex}\transpose \widehat{\bx}_{\otherSourceIndex} = 0$ for all $\sourceIndex \neq \otherSourceIndex$.
	\item[(a2)] $\bh_{\sourceIndex}\transpose \bh_{\otherSourceIndex} = 0$ or $ \widehat{\bh}_{\sourceIndex}\transpose \widehat{\bh}_{\otherSourceIndex} = 0$ for all $\sourceIndex \neq \otherSourceIndex$.
\end{description}
These orthogonality conditions are satisfied, e.g., when the vectors are sparse and their support does not overlap. 
In this way, two sparse inputs that are non-zero in non-overlapping nodes satisfy the first requirement in (a1). 
Similarly, two bandpass filters defined in non-overlapping bands satisfy the second requirement in (a2).

Before formally stating recoverability in the single graph case, we define auxiliary matrices to capture the orthogonality choices in (a1) and (a2). 
Consider the matrix $\bT_{\bx}=\bI$ if the first requirement in (a1) holds and $\bT_{\bx}=\bU$ if the second requirement is true. 
Similarly, consider $\bT_{\bh}=\bI$ if the first requirement in (a2) is true and $\bT_{\bh}=\bPsi$ if the second one holds. 
These matrices allow us to state recoverability conditions and a final algorithmic step for single graph demixing succinctly. 
That is, $\bT_{\bx}$ ($\bT_{\bh}$) captures whether the input signals (filter taps) are orthogonal in the node or spectral domains, rendering the remaining analysis independent of the domain.

\begin{myproposition}\label{P:SVD_source_sep}
	If the input signals $\{\bx_\sourceIndex^{(*)}\}_{\sourceIndex=1}^\numSources$ satisfy (a1), the diffusing graph filters $\{\bh_\sourceIndex^{(*)}\}_{\sourceIndex=1}^\numSources$ satisfy (a2) \textit{and} $\|\bT_{\bx}\bx_\sourceIndex^{(*)}\|\|\bT_{\bh}\bh_\sourceIndex^{(*)}\| \neq \|\bT_{\bx}\bx_{\otherSourceIndex}^{(*)}\|\|\bT_{\bh}\bh_{\otherSourceIndex}^{(*)}\|$ for all $\sourceIndex \neq \otherSourceIndex$; then $\{\bZ^{(*)}_\sourceIndex\}_{\sourceIndex=1}^\numSources$ can be recovered from $\bZ^{(*)}$.
\end{myproposition}

\begin{myproof}	
	First, form the matrix $\bT = \bT_{\bx} \bZ^{(*)} \bT_{\bh}\transpose$ and factorize it using the singular value decomposition (SVD) as $\bT = \bL \text{diag}(\boldsymbol{\sigma}) \bR\transpose$. 
	Then, the mixing components $\bZ_\sourceIndex^{(*)}$ are estimated individually by
	\begin{equation}\label{E:solution_from_svd} 
		\estimate{\bZ}_\sourceIndex^{(*)} = \sigma_\sourceIndex(\bT_{\bx}^{\dagger}) \bl_\sourceIndex \br_\sourceIndex\transpose (\bT_{\bh}^{\dagger})\transpose, 
	\end{equation}
	where $\sigma_\sourceIndex$ is the $\sourceIndex$th largest singular value of $\bT$, $\bl_\sourceIndex$ and $\br_\sourceIndex$ its associated left and right singular vectors, and $^\dagger$ denotes the pseudoinverse.
	
	The remaining part is to show that  $\estimate{\bZ}_\sourceIndex^{(*)} = \bZ_\sourceIndex^{(*)}$ for all $\sourceIndex$. 
	To this end, notice that we can write $\bT$ as 
	\begin{equation}\label{E:solution_from_svd_2} 
		\bT = \sum_{\sourceIndex}^{\numSources} \|\bT_{\bx}\bx_\sourceIndex^{(*)}\| \|\bT_{\bh}\bh_\sourceIndex^{(*)}\| \frac{\bT_{\bx}\bx_\sourceIndex^{(*)}}{\|\bT_{\bx}\bx_\sourceIndex^{(*)}\|} \frac{(\bT_{\bh}\bh_\sourceIndex^{(*)})\transpose}{\|\bT_{\bh}\bh_\sourceIndex^{(*)}\|}.
	\end{equation}
	From (a1)-(a2), it follows that ${\bT_{\bx}\bx_\sourceIndex^{(*)}}/{\|\bT_{\bx}\bx_\sourceIndex^{(*)}\|}$ for all $\sourceIndex$ forms an orthonormal basis and the same is true for ${\bT_{\bh}\bh_\sourceIndex^{(*)}}/{\|\bT_{\bh}\bh_\sourceIndex^{(*)}\|}$.
	Moreover, it is assumed in the statement of the proposition that all $\|\bT_{\bx}\bx_\sourceIndex^{(*)}\| \|\bT_{\bh}\bh_\sourceIndex^{(*)}\|$ are distinct.
	Thus, \eqref{E:solution_from_svd_2} must coincide with the SVD decomposition in that 
	\begin{equation}\label{E:solution_from_svd_3} 
		\bl_\sourceIndex \!= \!\frac{\bT_{\bx}\bx_\sourceIndex^{(*)}}{\|\bT_{\bx}\bx_\sourceIndex^{(*)}\|},\,\,\,\,\,
		\br_\sourceIndex \!= \!\frac{\bT_{\bh}\bh_\sourceIndex^{(*)}}{\|\bT_{\bh}\bh_\sourceIndex^{(*)}\|},\,\,\,\,\,
		\sigma_i \!= \!\|\bT_{\bx}\bx_\sourceIndex^{(*)}\| \|\bT_{\bh}\bh_\sourceIndex^{(*)}\|.
	\end{equation}
	From~\eqref{E:solution_from_svd_3} it readily follows that $\estimate{\bZ}_\sourceIndex^{(*)}$ in~\eqref{E:solution_from_svd} is equal to $\bx_\sourceIndex^{(*)} \bh_\sourceIndex^{(*)}{\transpose}$, as sought.
\end{myproof}

Proposition~\ref{P:SVD_source_sep} reveals that even for the highly challenging scenario of multiple signals being diffused and aggregated in a single graph, demixing is still possible under more stringent constraints. 
The analysis in this section also hints at the fact that the similarity between graphs might be an indicator of how hard the demixing problem is, with identical graphs being an extreme setting.
We revisit this point from an empirical viewpoint in the next section.

\vspace{1mm}
\noindent\textbf{Remark 2 (A priori information)} Blind demixing benefits when prior knowledge on $\{\bh_\sourceIndex^{(*)}, \bx_\sourceIndex^{(*)}\}_{\sourceIndex=1}^\numSources$ is available. Even though this applies for both demixing cases with multiple diffusing graphs or a single diffusing graph, a priori information is particularly helpful in the single-graph scenario, where the inherent ambiguity is larger. Two types of a priori information can be employed, \textit{probabilistic} and \textit{deterministic}. Probabilistic information can be incorporated by augmenting the objective with tractable (log) prior distributions. Alternatively, deterministic knowledge about entries of the signals $\bx_{\sourceIndex}^{(*)}$ or the filter taps $\bh_{\sourceIndex}^{(*)}$ can be incorporated in the form of linear equality constraints. In particular, if $\{\bh_\sourceIndex^{(*)}\}_{\sourceIndex=1}^\numSources$ is known then the structure of the optimization can be simplified. That is, the bilinearity disappears and the resultant problem can be handled using classical sparse signal reconstruction tools. \textcolor{blue}{Another case occurs when $\tK_\sourceIndex$ values of the signal $\bx_\sourceIndex^{(*)}$ are known}. Since we optimize over $\bZ_\sourceIndex$ instead of directly over $\bx_\sourceIndex$, exploiting this information requires considering the constraints
\begin{equation}
	\bZ_{\sourceIndex,\ell}\transpose \bx_{\sourceIndex,\ell+1}^{(*)} = \bZ_{\sourceIndex,\ell+1}\transpose \bx_{\sourceIndex,\ell}^{(*)} \quad \forall \ {\ell \in \mathcal{K}_\sourceIndex},
\end{equation}
where $\mathcal{K}_\sourceIndex=\{n_\ell\}_{\ell=1}^{\tK_\sourceIndex}$ collects the indexes of the entries of $\bx_\sourceIndex^{(*)}$ that are known. Note that the introduction of this type of constraints still renders a convex program.

\section{Numerical Results}\label{S:Simulations}

Five experiments are performed to illustrate and validate our proposed blind demixing approach. In the first two, we assess and compare the effectiveness of single-graph and multiple-graph demixing using synthetic random graphs. In the third experiment, we analyze multiple-graph demixing using real-world brain graphs (where nodes represent regions of the human brain and edges anatomical connections) associated with different individuals. In the fourth experiment, we analyze the recovery performance of single-graph demixing using a real-world social network and, in the last experiment, we rely on some of the graph-dependent parameters identified in Theorem~\ref{T:main_theorem} to predict demixing performance in real-world graphs. 

\textcolor{blue}{Throughout this section, we discuss the results from a collection of blind demixing problems, where we change the type of graphs, as well as the numbers of sources $\numSources$, filter coefficients $\numTaps_\sourceIndex$, and seeding nodes $\numNonzero_\sourceIndex$.}
The signal and filter estimates will be denoted by $\{\estimate{\bx}_\sourceIndex,\estimate{\bh}_\sourceIndex\}$. \textcolor{blue}{Unless otherwise stated, the GSO $\bS_\sourceIndex = \bA_\sourceIndex$ (the adjacency matrix), $\numTaps_\sourceIndex=\numTaps$; $\numNonzero_\sourceIndex=\numNonzero$ for all the sources; $\bx^{(*)}_\sourceIndex$ and $\bh^{(*)}_\sourceIndex$ are distributed according to standard multivariate Gaussians and rescaled to have unit norm.}
In the single-graph experiments, $\bx^{(*)}_\sourceIndex$ and $\bh^{(*)}_\sourceIndex$ are selected according to Proposition~\ref{P:SVD_source_sep}. That is, in addition to satisfying (a1) and (a2) in \Cref{SS:Ambiguity_single}, the vectors are normalized to have the same $\ell_1$ norm but different $\ell_2$ norm, so that it becomes possible to resolve the ambiguity intrinsic to single-graph demixing.
Figures of merit derived from the demixing error (DE)
\begin{equation}
	\text{DE}:=\frac{1}{\numSources}\sum_{\sourceIndex=1}^\numSources\left\|\estimate{\bx}_\sourceIndex\estimate{\bh}_\sourceIndex\transpose-
	\bx^{(*)}_\sourceIndex{\bh^{(*)}_\sourceIndex}\transpose\right\|_\text{F},
\end{equation}
are adopted to assess demixing. These are the median DE across realizations and the \emph{rate of successful recovery} defined to be the probability that $\text{DE}<10^{-3}$. \textcolor{blue}{Recall that in Section~\ref{S:BlindDem} we highlighted that the problem is bilinear and thus poses a scaling ambiguity between the identified filters and separated signals. Employing the DE deals effectively with this ambiguity during the performance assessment.} The results reported correspond to $1000$ demixing realizations each of them solved using the log-det relaxation (see Section \ref{SS:Convex_relax_multiple} and \cite{Fazel2003,ramirez2017graph} for additional details).

\noindent\textbf{Blind demixing in a single random graph.} \Cref{fig:singlegraph} shows the results of demixing with a single diffusing graph of varying $\numNodes$ and $\numTaps=3$. Random graphs are drawn according to the Erd\H{o}s-R\'enyi model \cite{bollobas1998random} with edge probability $0.1$. Four settings with different values of $\numSources$ and $\numNonzero$ are presented. As expected, the performance worsens for larger values of $\numSources$ and $\numNonzero$. Interestingly, demixing for $(\numSources\!=\!3, \numNonzero\!=\!6)$ is less successful than for $(\numSources\!=\!2, \numNonzero\!=\!9)$. Since the number of signal coefficients to estimate in these two settings is the same ($\numSources\numNonzero$), this suggests that the recovery is more sensitive to $\numSources$.

\noindent\textbf{Effect of graph similarity on recovery.} The second experiment explores demixing with two Barab\'asi-Albert graphs \cite{albert2002statistical} of varying $\numNodes$ (built following the scale-free model from a seeding Erd\H{o}s-R\'enyi graph of $\numNodes/10$ nodes and edge probability $0.1$) and $\numTaps=3$. The two graphs are generated so that $\alpha$ percent of their edges are the same. Note that $\alpha=1$ leads to demixing with a single diffusing graph. \Cref{fig:twograph} shows that, as expected, the multi-graph demixing strategy (cf. \Cref{S:BlindDem}) fails for $\alpha=1$, although the performance does not considerably drop until $\alpha>0.9$.

\noindent\textbf{Blind demixing with multiple brain graphs.} {\Cref{fig:brain} depicts the rate of successful recovery using up to five different \emph{brain graphs} corresponding to different people. The graphs belong to a dataset of six undirected graphs of the human brain, consisting of $\numNodes=66$ regions of interest (ROI) and edges given by the density of anatomical connections between regions~\cite{hagmann2008mapping}}. The level of activity of each ROI in a brain graph can be represented by a graph signal $\bx_\sourceIndex$, thus, successive applications of $\bS_\sourceIndex$ model a linear evolution of the brain activity pattern. Under the hypothesis that we observe a sum of linear combinations (filtered signals) resulting from diffusing originally sparse brain signals, then blind demixing jointly estimates which regions of the brains were originally active, the activity in these regions, and the diffusing coefficients $\bh_\sourceIndex$ of the linear combinations. In this experiment, $\numNonzero=1$ while $\numTaps$ and $\numSources$ vary. The results show that demixing is indeed feasible, although the performance decreases quickly as $\numSources$ increase. This is not surprising since the brain graphs at hand exhibit strong similarities. On the other hand, for a fixed value of $\numSources$ it is interesting that the performance increases with $\numTaps$ growing from 2 to 4, and it starts to drop afterwards. This reveals a trade-off between the number of parameters to estimate (increasing with $\numTaps$) and the amount of information of $\bx_{\sourceIndex}$ in $\by$ (which increases with the successive diffusions).

\begin{figure}
	\hspace{-5mm}
	\includegraphics[width=0.55\textwidth]{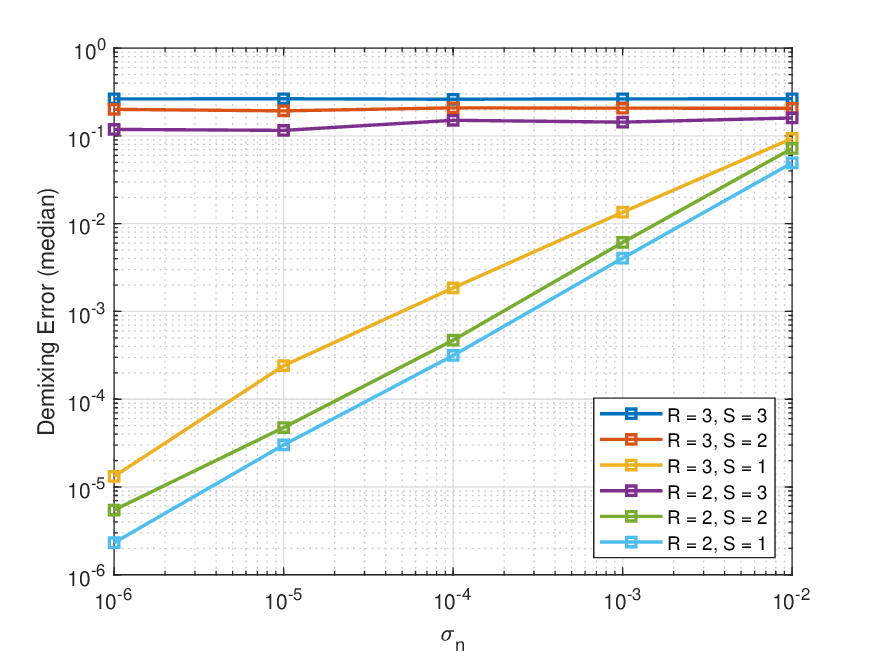}
	\caption{ \small Blind demixing in the karate network. Demixing Error (DE) as a function of the noise level $\sigma_n$ for different number of filters ($\numSources$) and seeding nodes ($\numNonzero$). The DE decreases as the sum $\numSources+\numNonzero$ also decreases. When $\numSources+\numNonzero$ is fixed, smaller $\numSources$ results in higher demixing accuracy. {In the three cases with $\numSources+\numNonzero\ge5$ where the error does not converge to zero as the noise power goes to zero, demixing is too  hard given the size of the karate network ($\numNodes=34$) to achieve successful recovery.}}
	\label{Fig:singlegraph_karate}
\end{figure}

\noindent\textbf{Effect of noise on recovery.} In this experiment, we study the performance of single-graph demixing \eqref{eq:convex_minization_sum} with $\numTaps=3$, varying $\numSources$ and $\numNonzero$, and in the presence of additive white Gaussian noise $\mathcal{N}(0,\sigma_n^2)$ (see \Cref{remark:robust_demixing} for a discussion on handling noisy observations). The graph at hand corresponds to the adjacency matrix of Zachary's karate club network~\cite{zachary1977information}, which contains $\numNodes=34$ nodes (the members of the club) and $78$ undirected edges (symbolizing friendships among members). The results reported in \Cref{Fig:singlegraph_karate} reveal that, in all cases, the demixing error decreases as $\sigma_n$ decreases and it increases as either $\numSources$ or $\numNonzero$ increase. Furthermore, if the value $\numSources+\numNonzero$ is fixed, successful demixing is easier for smaller values of $\numSources$. Once again, this suggests that the sensitivity of the demixing performance with respect to $\numSources$ is larger than that for $\numNonzero$.

\noindent\textbf{Predicting demixing performance for brain graphs.} In Section \ref{S:Recovery}, we presented a theoretical result characterizing demixing performance as a function of a number of parameters. When taking a close look at the proof (\Cref{sec:coherence_pairs}), we observe that the values of $\rho_{\bU_{\yetOtherSourceIndex}}^{(\numNonzero_\yetOtherSourceIndex)}$, $\rho_{\bU_{\otherSourceIndex}}^{(\numNonzero_\otherSourceIndex)}$,  $\rho_{\bPsi_{\yetOtherSourceIndex}}^{(\numTaps_\yetOtherSourceIndex)}$ and $\rho_{\bPsi_{\otherSourceIndex}}^{(\numTaps_\otherSourceIndex)}$  are intrinsic to demixing, account for the interaction between mixing components, and have a larger impact on recovery (see \Cref{sec:union_bound} for details).

This motivates us to combine the values of $\rho_{\bU_{\yetOtherSourceIndex}}^{(\numNonzero_\yetOtherSourceIndex)}$, $\rho_{\bU_{\otherSourceIndex}}^{(\numNonzero_\otherSourceIndex)}$,  $\rho_{\bPsi_{\yetOtherSourceIndex}}^{(\numTaps_\yetOtherSourceIndex)}$ and $\rho_{\bPsi_{\otherSourceIndex}}^{(\numTaps_\otherSourceIndex)}$ using the simple expression
\begin{equation}
	\label{eq:brain_predictor}
	\bar{\rho}_{\yetOtherSourceIndex,\otherSourceIndex}:=\rho_{\bU_{\yetOtherSourceIndex}}^{(\numNonzero_\yetOtherSourceIndex)} \ \rho_{\bU_{\otherSourceIndex}}^{(\numNonzero_\otherSourceIndex)} \ \rho_{\bPsi_{\yetOtherSourceIndex}}^{(\numTaps_\yetOtherSourceIndex)} \ \rho_{\bPsi_{\otherSourceIndex}}^{(\numTaps_\otherSourceIndex)},
\end{equation}
and, then, employ $\bar{\rho}_{\yetOtherSourceIndex,\otherSourceIndex}$ as a predictor of the demixing performance between a pair of graphs. 
Specifically, these quantities stem from the center parameter of the concentration inequality (i.e. the bound in \eqref{eq:coherence_concentration_inequality_center}) and the relation between $\kappa$ and $\rho$ in~\eqref{eq:kappa_rho_relation}.

Hence, the results in this experiment try to asses the dependence of the demixing performance with respect to $\bar{\rho}_{\yetOtherSourceIndex,\otherSourceIndex}$. To that end, we consider the six brain graphs used in the second experiment (cf.~\cite{hagmann2008mapping}), and solve 15 (six choose two) different demixing problems, each of them involving two out of the six graphs with $\numTaps=2$ and $\numNonzero=1$. For each pair of graphs, we use the Laplacian as GSO, compute $\bar{\rho}_{\yetOtherSourceIndex,\otherSourceIndex}$ according to \eqref{eq:brain_predictor} and quantify the rate of successful recovery.

%
%


\begin{figure}
	\centering
	
	\begin{subfigure}{0.4\textwidth}
		\centering
		\includegraphics[width=\textwidth]{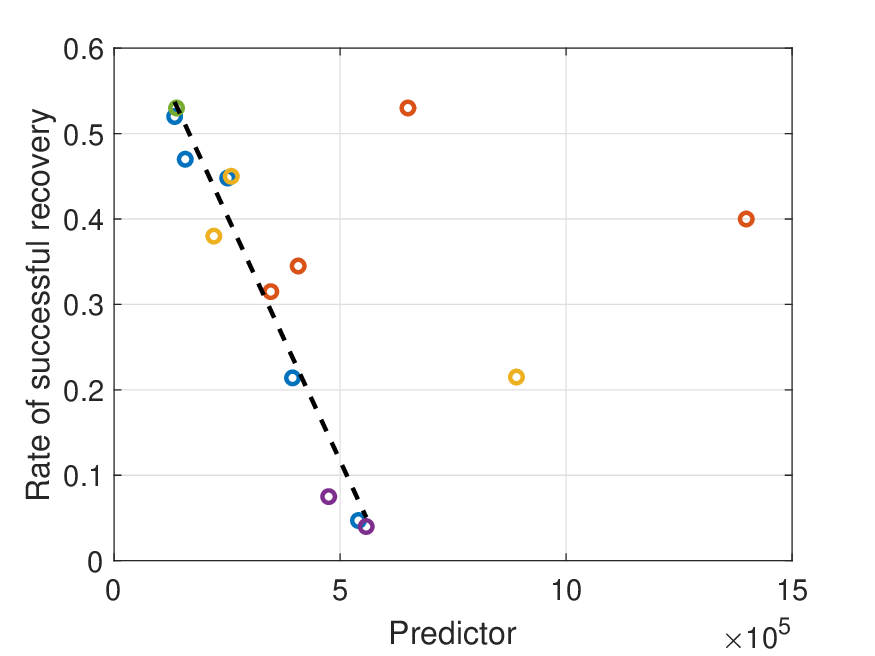}
		\caption{}
		\label{fig:predictor_results}
	\end{subfigure} \\
	\begin{subfigure}{0.4\textwidth}
		\centering\includegraphics[width=\textwidth]{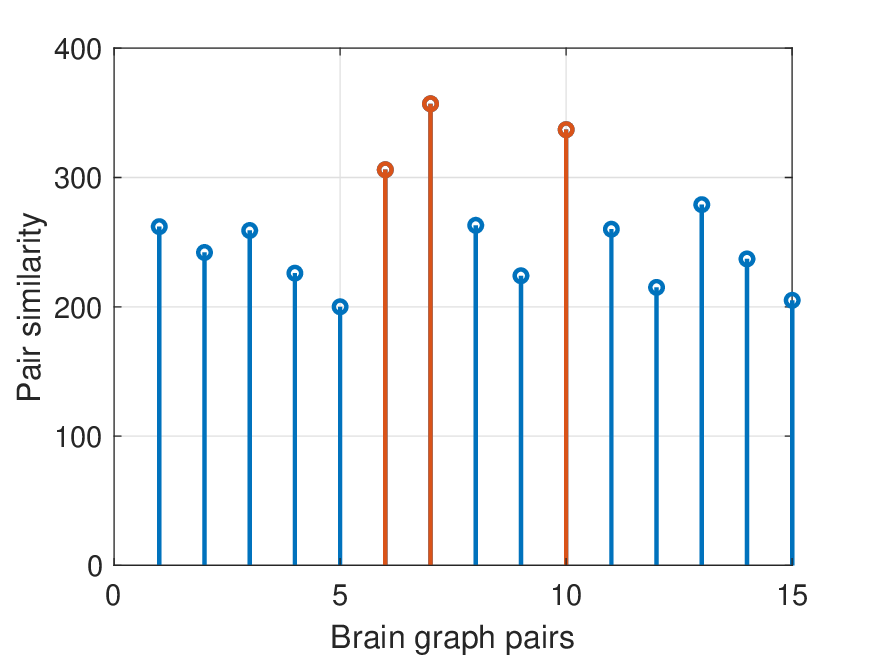}
		\caption{}
		\label{fig:predictor_explain}
	\end{subfigure}
	
	\vspace{-0.1in}
	\caption{\small (a) Recovery rate of the convex blind demixing minimization [cf.~\eqref{eq:convex_minization2}] as a function of the predictor~\eqref{eq:brain_predictor}. \textcolor{blue}{Each point corresponds to a pair of brain graphs, where points of the same color share one fixed brain graph (of the six in total).} There are three outliers on the right side of the plot. The straight line, result of a robust linear regression, shows the correlation between the predictor and the recovery rate. (b) Explanation of the outliers in (a) based on graph similarity for all pairs of brain graphs. The three peaks correspond to the outliers in (a).}
\end{figure}

The obtained results, which are reported in \Cref{fig:predictor_results}, confirm that pairs of graphs with a larger value of $\bar{\rho}_{\yetOtherSourceIndex,\otherSourceIndex}$ give rise to more challenging demixing scenarios (i.e., smaller rate of recovery). In fact, if we fit the obtained data using a linear model robust against outliers, we obtain the dashed-straight line shown in the figure. This simple analysis reveals that most of the points fall in the close vicinity of the linear fit, pointing out that the predictor in  \eqref{eq:brain_predictor} is a good \emph{graph-dependent predictor} for demixing performance and, implicitly, validating the analysis and assumptions in Theorem~\ref{T:main_theorem}. Interestingly, the three pairs of graphs that are deemed as outliers are the ones with most overlapping edges, as illustrated in \Cref{fig:predictor_explain}, which reports the number of overlapping edges for each pair of graphs and highlights the three outliers identified in \Cref{fig:predictor_results}. This could imply that some of the assumptions adopted during the proof of the theorem could be too pessimistic when the graphs at hand are very similar, opening the door to future work focused on those types of graphs.


\section{Conclusions}\label{S:Conclusions}

We formulated and analyzed the problem of blind demixing of filtered diffused graph signals, an extension of blind demixing of time (or spatial) domain signals to graphs. Envisioned application domains encompass a particular class of bilinear inverse problems, where the observed graph signal is modeled as a linear combination of  diffusion processes driven by a number of sources. 
While the {graph signal} observations are bilinear functions of the filter coefficients and the sparse input signal, we rely on the frequency interpretation of graph signals and filters to show that they are also linearly related to the entries of a lifted rank-one, row-sparse matrix. Accordingly, the blind demixing problem of graph signals was tackled via rank and sparsity minimization subject to linear constraints, an  inverse problem amenable to convex relaxations offering provable recovery guarantees under simplifying assumptions.
The theoretical results reveal key features driving the demixing performance including the number of sources and graph nodes, the sparsity levels, the filter orders, as well as spectral properties of the underlying graphs and their {pairwise coherences}. These spectral properties are captured via conveniently defined {concentration} parameters.
Numerical tests validated the theoretical claims and demonstrated that the proposed approach offers satisfactory recovery performance, even for settings well beyond the scope of the analysis.
Building on the insights gained here, it is of interest to develop spectral graph theoretical tools to facilitate identifying the key relationships between the (e.g., topological) properties of graphs and the resulting recovery performance. Future research includes also estimating the shift operator employing network topology inference~\cite{mei2016signal,segarra2017network,zhu2020network}.

\appendix

\subsection{Proof of Theorem~\ref{T:main_theorem}}\label{App:proof_the_noise_free}

The proof is based on a set of conditions outlined in {Lemma}~\autoref{lemma:sufficient_conditions}. In~\Cref{sec:suff_conditions} we show that, under these conditions, the nuclear norm minimization \eqref{eq:nuclear_optimization} produces the same result as the original constrained search~\eqref{eq:feasibility2}. 
For the remaining of the proof, \Cref{sec:coherence_pairs,sec:dual_certificate} establish the conditions and, finally, \Cref{sec:union_bound} concludes the proof by implementing a relevant union bound.
We begin by introducing notation that will be used throughout the proof.

Without loss of generality, it is assumed throughout that the true input signals \textcolor{blue}{$\bs_\sourceIndex^{(*)}, \bx_\sourceIndex^{(*)}$} and filter taps $\bh_\sourceIndex^{(*)}$ are real vectors with unit norm.
\textcolor{blue}{Subsequently, with $\bZ_\sourceIndex \in \mathbb{C}^{\numNonzero_\sourceIndex \times \numTaps_\sourceIndex}$, regard the linear subspaces $\mT_\sourceIndex$ wherein $\bs_\sourceIndex^{(*)} \bh_\sourceIndex^{(*)}{\transpose}$ lies, as well as their complements $\mT_\sourceIndex^\perp$,}
\textcolor{blue}{
\begin{align}
	\mT_\sourceIndex = \{\bZ_\sourceIndex \bh_\sourceIndex^{(*)} \bh_\sourceIndex^{(*)}{\transpose} \! + \bs_\sourceIndex^{(*)} \bs_\sourceIndex^{(*)}{\transpose} \bZ_\sourceIndex (\bI_{\numTaps_\sourceIndex} - \bh_\sourceIndex^{(*)} \bh_\sourceIndex^{(*)}{\transpose})\}, \\
	\mT_\sourceIndex^\perp = \{(\bI_{\numNonzero_\sourceIndex} - \bs_\sourceIndex^{(*)} \bs_\sourceIndex^{(*)}{\transpose}) \bZ_\sourceIndex (\bI_{\numTaps_\sourceIndex} - \bh_\sourceIndex^{(*)} \bh_\sourceIndex^{(*)}{\transpose})\},
\end{align}
}
\hspace{-2mm} Furthermore, let \textcolor{blue}{$\ba_{\sourceIndex,\nodeIndex}\transpose$, $\bu_{\sourceIndex,\nodeIndex}\transpose$} and $\bpsi_{\sourceIndex,\nodeIndex}\transpose$ denote the $\nodeIndex$th row of the GFT matrices \textcolor{blue}{$\bA_\sourceIndex$, $\bU_\sourceIndex$} and $\bPsi_\sourceIndex$, respectively (cf. \Cref{s:gsp_frequency}). 
Then, consider the following linear operators \textcolor{blue}{$\decoupledmodelop_\sourceIndex: \mathbb{C}^{\numNonzero_\sourceIndex \times \numTaps_\sourceIndex} \to \mathbb{C}^{\numNodes}$,} $\modelop_\sourceIndex: \mathbb{C}^{\numNodes \times \numTaps_\sourceIndex} \to \mathbb{C}^{\numNodes}$ and their adjoints
\textcolor{blue}{$\decoupledmodelop^*_\sourceIndex: \mathbb{C}^{\numNodes} \to \mathbb{C}^{\numNonzero_\sourceIndex \times \numTaps_\sourceIndex}$,} $\modelop^*_\sourceIndex: \mathbb{C}^{\numNodes} \to \mathbb{C}^{\numNodes \times \numTaps_\sourceIndex}$,
\textcolor{blue}{
\begin{align}
	\decoupledmodelop_\sourceIndex(\bZ) &= \left\lbrace \ba_{\sourceIndex,\nodeIndex}\transpose \bZ \bpsi_{\sourceIndex,\nodeIndex} \right\rbrace_{\nodeIndex=1}^\numNodes, \quad \bZ \in \mathbb{C}^{\numNonzero_\sourceIndex \times \numTaps_\sourceIndex} \\
	\decoupledmodelop^*_\sourceIndex(\bz) &= \sum_{\nodeIndex=1}^\numNodes z_\nodeIndex \overline{\ba_{\sourceIndex,\nodeIndex}} \bpsi_{\sourceIndex,\nodeIndex}\hermitian, \quad \bz \in \mathbb{C}^{\numNodes} 
\end{align}
}
\begin{align}
	\modelop_\sourceIndex(\bZ) &= \left\lbrace \bu_{\sourceIndex,\nodeIndex}\transpose \bZ \bpsi_{\sourceIndex,\nodeIndex} \right\rbrace_{\nodeIndex=1}^\numNodes, \quad \bZ \in \mathbb{C}^{\numNodes \times \numTaps_\sourceIndex}, \\
	\modelop^*_\sourceIndex(\bz) &= \sum_{\nodeIndex=1}^\numNodes z_\nodeIndex \overline{\bu_{\sourceIndex,\nodeIndex}} \bpsi_{\sourceIndex,\nodeIndex}\hermitian, \quad \bz \in \mathbb{C}^{\numNodes}.
\end{align}
\textcolor{blue}{In the sequel, we employ conditions with $\modelop_\sourceIndex$ restricted to $\Omega_\sourceIndex$ in order to prove conditions with $\decoupledmodelop_\sourceIndex$ restricted to $\mT_\sourceIndex$.}

Let us also introduce a \emph{common partition} $\{\partion_\partionIndex\}_{\partionIndex=1}^\partionSize$ of $\{1,2,\dots,\numNodes\}$ with $|\partion_\partionIndex|=\numPartionElems \ \forall \ \partionIndex \in \{1,2,\dots,\partionSize\}$ and $\numNodes=\partionSize \numPartionElems$. See \cite[section II-C]{ling2017blind} and references therein for further discussion on the partitioning. Associated with the partition, define the matrices
\begin{equation}
\label{eq:partion_matrices}
\partmat_{\sourceIndex,\partionIndex} = \sum_{\ell \in \partion_\partionIndex} \bpsi_{\sourceIndex,\ell} \bpsi_{\sourceIndex,\ell}\hermitian,  \qquad \parimat_{\sourceIndex,\partionIndex} = \partmat_{\sourceIndex,\partionIndex}^{-1},
\end{equation}
the sampled linear operators
\begin{align*}
	\modelop_{\sourceIndex,\partionIndex}(\bZ) &= \{ \bu_{\sourceIndex,\ell}\transpose \bZ \bpsi_{\sourceIndex,\ell} \}_{\ell \in \partion_\partionIndex}, \\
	\modelop^*_{\sourceIndex,\partionIndex}(\bz) &= \sum_{\ell \in \partion_\partionIndex} z_\ell \overline{\bu_{\sourceIndex,\ell}} \bpsi_{\sourceIndex,\ell}\hermitian,
\end{align*}
and the following scalar quantities
\begin{equation}
	\label{eq:muh}
	\mu_h^2 := \max \Bigg\{ \frac{\partionSize^2}{\numNodes} \max_{ \substack{\ell \in \partion_\partionIndex, \\ 1 \le \partionIndex \le \partionSize, \\ 1 \le \sourceIndex \le \numSources}} \frac{|\bpsi_{\sourceIndex,\ell}\hermitian \parimat_{\sourceIndex,\partionIndex} \bh_\sourceIndex|^2}{\|\bh_\sourceIndex\|^2}, \numNodes \max_{\substack{1 \le \nodeIndex \le \numNodes, \\ 1 \le \sourceIndex \le \numSources}} \frac{|\bpsi_{\sourceIndex,\nodeIndex}\hermitian \bh_\sourceIndex|^2}{\|\bh_\sourceIndex\|^2} \Bigg\}
\end{equation}
%
%
\begin{equation}
	\label{eq:mumax}
	\mu_{\max}^2 := \max_{1 \le \nodeIndex \le \numNodes, 1 \le \sourceIndex \le \numSources} \frac{\numNodes}{\numTaps_\sourceIndex} \|\bpsi_{\sourceIndex, \nodeIndex}\|^2,
\end{equation}
%

Additionally, we will handle projections on spaces such as those constrained to the signals supports $\mS_\sourceIndex$,
\begin{align*}
{\modelop_{\sourceIndex}}_{\mS_\sourceIndex}(\bZ) &= \{ {\bu_{\sourceIndex,\nodeIndex}\transpose}_{\mS_\sourceIndex} \bZ \bpsi_{\sourceIndex,\nodeIndex} \}_{\nodeIndex=1}^\numNodes, \\
{\modelop^*_{\sourceIndex}}_{\mS_\sourceIndex}(\bz) &= \sum_{\nodeIndex=1}^\numNodes z_\nodeIndex {\overline{\bu_{\sourceIndex,\nodeIndex}}}_{\mS_\sourceIndex} \bpsi_{\sourceIndex,\nodeIndex}\hermitian.
\end{align*}
Naturally, projected \emph{and} sampled operators will be denoted by
\begin{align*}
{\modelop_{\sourceIndex,\partionIndex}}_{\mS_\sourceIndex}(\bZ) &= \{ {\bu_{\sourceIndex,\ell}\transpose}_{\mS_\sourceIndex} \bZ \bpsi_{\sourceIndex,\ell} \}_{\ell\in\partion_\partionIndex}, \\
{\modelop^*_{\sourceIndex,\partionIndex}}_{\mS_\sourceIndex}(\bz) &= \sum_{\ell \in \partion_\partionIndex} z_\ell {\overline{\bu_{\sourceIndex,\ell}}}_{\mS_\sourceIndex} \bpsi_{\sourceIndex,\ell}\hermitian.
\end{align*}
Although making slight abuse overloading the symbols, {the subindex notation denoting projection onto a space} will be used indistinctly for linear operators, matrices, and vectors. \textcolor{blue}{Finally, when the projection notation via subindex is not suitable, the linear operator $\project_{\mT}$ will be used instead to denote the projection onto a space $\mT$.}

\subsubsection{Sufficient conditions}
\label{sec:suff_conditions}

In \cite[Lemma 5.1]{ling2017blind} it is shown that for any real $\{\bH_\sourceIndex\}_{\sourceIndex=1}^\numSources$ satisfying that \textcolor{blue}{$\sum_{\sourceIndex=1}^\numSources \decoupledmodelop_\sourceIndex(\bH_\sourceIndex)=0$} with at least one $\bH_\sourceIndex$ nonzero, if
%
%
\begin{equation}
	\label{eq:lemma51_ling2017blind}
	\textcolor{blue}{\sum_{\sourceIndex=1}^\numSources \left\langle\bH_\sourceIndex, \bs_\sourceIndex^{(*)}\bh_\sourceIndex^{(*)}{\transpose}\right\rangle + \|\bH_{\sourceIndex,\mT_\sourceIndex^\perp}\|_* > 0}
\end{equation}
then the solution to the convex program,
%
%
\begin{equation}\label{E:convex_program_freq}
	\textcolor{blue}{\min_{\{\bs_\sourceIndex,\bh_\sourceIndex\}_{\sourceIndex=1}^\numSources} \sum_{\sourceIndex=1}^{\numSources} \|\bs_\sourceIndex \bh_\sourceIndex\transpose\|_* \quad \text{s. to} \quad \widehat{\bnu} = \sum_{\sourceIndex=1}^{\numSources} \decoupledmodelop_\sourceIndex\left(\bs_\sourceIndex \bh_\sourceIndex\transpose\right),}
\end{equation}
is the set formed by the true \textcolor{blue}{$\bs_\sourceIndex^{(*)}$} and $\bh_\sourceIndex^{(*)}$. Nevertheless, this lemma does not provide with conditions on the linear operators, which capture the blind demixing model. 
To this end, inspired by~\cite[{Lemma 5.2}]{ling2017blind}, we state the following lemma.

\begin{mylemma}\label{lemma:sufficient_conditions}
	Assume that, for all $\sourceIndex$,
	%
	%
	\begin{equation}
		\label{eq:cond_local_isometry}
		\textcolor{blue}{\|\project_{\mT	_\sourceIndex}\decoupledmodelop_\sourceIndex^*\decoupledmodelop_\sourceIndex\project_{\mT_\sourceIndex} - \project_{\mT_\sourceIndex}\| \le \frac{1}{4},}
	\end{equation}
	%
	%
	\begin{equation}
		\label{eq:cond_cross_coherence}
		\textcolor{blue}{\mu := \max_{\otherSourceIndex \neq \yetOtherSourceIndex}  \|\project_{\mT_\otherSourceIndex}\decoupledmodelop_\otherSourceIndex^*\decoupledmodelop_\yetOtherSourceIndex\project_{\mT_\yetOtherSourceIndex}\| \le \frac{1}{4\numSources},}
	\end{equation}
	%
     %
	\begin{equation}
		\label{eq:cond_operator_norm}
		\textcolor{blue}{\|\decoupledmodelop_\sourceIndex\| \le \gamma,}
    \end{equation}
	for some $\gamma>0$, and that there exists a $\blambda \in \mathbb{C}^\numNodes$ (denominated \emph{dual certificate}) such that, for all $i$,
	%
	%
	\begin{equation}
		\label{eq:cond_dual0_sparse}
		\textcolor{blue}{{\| \bs_\sourceIndex^{(*)}\bh_\sourceIndex^{(*)}{\transpose} - \project_{\mT_\sourceIndex}({\decoupledmodelop_\sourceIndex^*}(\blambda)) \|_{\text{F}} \le \frac{1}{5 \numSources \gamma}},}
	\end{equation}
	%
	%
	\begin{equation}
		\label{eq:cond_dual1_sparse}
		\textcolor{blue}{\| \project_{\mT_\sourceIndex^\perp}(\decoupledmodelop_\sourceIndex^*(\blambda)) \| \le \frac{1}{2}.}
	\end{equation}
	Then, the solution to~\eqref{E:convex_program_freq} is given by the true signals \textcolor{blue}{$\bs_\sourceIndex^{(*)}$} and filter taps $\bh_\sourceIndex^{(*)}$.
\end{mylemma}

\begin{myproof}
	We begin by defining $I_1$,
	\begin{equation*}
		I_1 := \sum_{\sourceIndex=1}^\numSources \left\langle\bH_\sourceIndex,\textcolor{blue}{\bs}_\sourceIndex^{(*)}\bh_\sourceIndex^{(*)}{\transpose}-\textcolor{blue}{\decoupledmodelop}_\sourceIndex^*(\blambda)\right\rangle + \|\bH_{\sourceIndex,\textcolor{blue}{\mT}_\sourceIndex^\perp}\|_*.
	\end{equation*}
	Our goal is to show that for any nonzero $\{\bH_\sourceIndex\}_{\sourceIndex=1}^\numSources$ such that $\sum_{\sourceIndex=1}^\numSources \textcolor{blue}{\decoupledmodelop}_\sourceIndex(\bH_\sourceIndex)=0$, it holds that $I_1 > 0$. The inner product in $I_1$ is decomposed into \textcolor{blue}{$\mT_\sourceIndex$ and $\mT_\sourceIndex^\perp$} to obtain
	\begin{equation*}
		\begin{split}
			I_1 = \sum_{\sourceIndex=1}^\numSources &\ \Big(\langle\bH_{\sourceIndex,\textcolor{blue}{\mT}_\sourceIndex}, \textcolor{blue}{\bs}_\sourceIndex^{(*)}\bh_\sourceIndex^{(*)}{\transpose}-\textcolor{blue}{\project_{\mT_\sourceIndex}}(\textcolor{blue}{\decoupledmodelop}_\sourceIndex^*(\blambda)) \rangle \\
			&-\langle\bH_{\sourceIndex,\textcolor{blue}{\mT}_\sourceIndex^\perp}, \textcolor{blue}{\project_{\mT_\sourceIndex^\perp}}(\textcolor{blue}{\decoupledmodelop}_\sourceIndex^*(\blambda)) \rangle + \|\bH_{\sourceIndex,\textcolor{blue}{\mT}_\sourceIndex^\perp}\|_*\Big).
		\end{split}
	\end{equation*}
	Let us additionally define $I_2$ as follows
	\begin{equation*}
		\begin{split}
			I_2 := \sum_{\sourceIndex=1}^\numSources \Big( -&\|\bH_{\sourceIndex,\textcolor{blue}{\mT}_\sourceIndex}\|_\tF \ \|\textcolor{blue}{\bs}_\sourceIndex^{(*)}\bh_\sourceIndex^{(*)}{\transpose}-\textcolor{blue}{\project_{\mT_\sourceIndex}(}\textcolor{blue}{\decoupledmodelop}_\sourceIndex^*(\blambda)\textcolor{blue}{)}\|_\tF \\ +
			&\|\bH_{\sourceIndex,\textcolor{blue}{\mT}_\sourceIndex^\perp}\|_* (1-\|\textcolor{blue}{\project_{\mT_\sourceIndex^\perp}(}\textcolor{blue}{\decoupledmodelop}_\sourceIndex^*(\blambda)\textcolor{blue}{)}\|)\Big).
		\end{split}
	\end{equation*}

	We are now going to rely on the following two matrix inequalities. First,
	\begin{equation}\label{E:matrix_ineq_1}
		\langle\bU, \bV\rangle \ge -|\langle\bU, \bV\rangle| \ge -\|\bU\|_\tF \|\bV\|_\tF,
	\end{equation}
	which follows from the Cauchy-Schwarz inequality. Second,
	\begin{equation}\label{E:matrix_ineq_2}
		-\langle\bU, \bV\rangle \ge -|\langle\bU, \bV\rangle| \ge -\|\bU\|_* \|\bV\|,
	\end{equation}
	which can be shown \textcolor{blue}{by applying Cauchy-Schwarz and the definitions of the $\ell_2$ and the nuclear norms of matrices.}
	Leveraging~\eqref{E:matrix_ineq_1} and~\eqref{E:matrix_ineq_2}, it follows that $I_1 \ge I_2$. Thus, we now seek to show that $I_2 > 0$.

    Making use of $\|\cdot\|_* \ge \|\cdot\|_\tF$ we have that
	\textcolor{blue}{
	\begin{equation*}
		\begin{split}
			I_2 \ge \sum_{\sourceIndex=1}^\numSources \Big(&-\|\bH_{\sourceIndex,\mT_\sourceIndex}\|_\tF \|\bs_\sourceIndex^{(*)}\bh_\sourceIndex^{(*)}{\transpose}-\project_{\mT_\sourceIndex}(\decoupledmodelop_\sourceIndex^*(\blambda))\|_\tF \\
			&+\|\bH_{\sourceIndex,\mT_\sourceIndex^\perp}\|_\tF(1-\|\project_{\mT_\sourceIndex^\perp}(\decoupledmodelop_\sourceIndex^*(\blambda))\|)\Big).
		\end{split}
	\end{equation*}
	}
	Next, {from \eqref{eq:cond_dual0_sparse} and \eqref{eq:cond_dual1_sparse}} it follows that
	\textcolor{blue}{
	\begin{equation*}
	I_2 \ge -\frac{1}{5\numSources\gamma}\sum_{\sourceIndex=1}^\numSources \|\bH_{\sourceIndex,\mT_\sourceIndex}\|_\tF + \frac{1}{2}\sum_{\sourceIndex=1}^\numSources \|\bH_{\sourceIndex,\mT_\sourceIndex^\perp}\|_\tF.
	\end{equation*}
	}
	Moreover, from {\eqref{eq:cond_local_isometry}-\eqref{eq:cond_operator_norm}}, it can be shown that
	\textcolor{blue}{
	\begin{equation}
		\label{eq:sufficient_conditions_ingredient}
		\frac{1}{2 \numSources} \sum_{\sourceIndex=1}^\numSources \|\bH_{\sourceIndex,\mT_\sourceIndex}\|_\tF \le \gamma \sum_{\sourceIndex=1}^\numSources \|\bH_{\sourceIndex,\mT_\sourceIndex^\perp}\|_\tF.
	\end{equation}
	}
	The derivation of~\eqref{eq:sufficient_conditions_ingredient} follows the same steps as those in the proof of~\cite[{Lemma 5.2}]{ling2017blind}. Hence, we avoid it here to reduce redundancy.
	
	Leveraging~\eqref{eq:sufficient_conditions_ingredient} we obtain that
	\textcolor{blue}{
	\begin{equation*}
		I_2 \ge -\frac{2}{5}\sum_{\sourceIndex=1}^\numSources \|\bH_{\sourceIndex,\mT_\sourceIndex^\perp}\|_\tF + \frac{1}{2}\sum_{\sourceIndex=1}^\numSources \|\bH_{\sourceIndex,\mT_\sourceIndex^\perp}\|_\tF = \frac{1}{10}\sum_{\sourceIndex=1}^\numSources\|\bH_{\sourceIndex,\mT_\sourceIndex^\perp}\|_\tF,
	\end{equation*}
	}
	which is indeed larger than zero since at least one of the terms $\|\bH_{\sourceIndex,\textcolor{blue}{\mT}_\sourceIndex^\perp}\|_\tF$ in the sum must be nonzero owing to the fact that via \eqref{eq:sufficient_conditions_ingredient} $\bH_{\sourceIndex,\textcolor{blue}{\mT}_\sourceIndex^\perp}\!=\!0, \ 1 \! \le \! \sourceIndex  \! \le \! \numSources$ would imply that all $\bH_{\sourceIndex}\!=\!0$.
	
	Finally, knowing that $I_2>0$ and therefore $I_1>0$ too, from the definition of an adjoint operator it follows that
	\begin{align*}
	I_1 & = \sum_{\sourceIndex=1}^\numSources \left\langle\bH_\sourceIndex, \textcolor{blue}{\bs}_\sourceIndex^{(*)}\bh_\sourceIndex^{(*)}{\transpose}\right\rangle 
	- \sum_{\sourceIndex=1}^\numSources \left\langle \textcolor{blue}{\decoupledmodelop}_\sourceIndex(\bH_\sourceIndex), \blambda \right\rangle 
	+ \|\bH_{\sourceIndex,\textcolor{blue}{\mT}_\sourceIndex^\perp}\|_*\nonumber \\
	& = \sum_{\sourceIndex=1}^\numSources \left\langle\bH_\sourceIndex, \textcolor{blue}{\bs}_\sourceIndex^{(*)}\bh_\sourceIndex^{(*)}{\transpose}\right\rangle 
	+ \|\bH_{\sourceIndex,\textcolor{blue}{\mT}_\sourceIndex^\perp}\|_* > 0.
	\end{align*}
	Having shown~\eqref{eq:lemma51_ling2017blind}, the statement of the lemma holds.
\end{myproof}

\vspace{1.5mm}

At this point we are left occupied with showing~\eqref{eq:cond_local_isometry}-\eqref{eq:cond_dual1_sparse}. 
\textcolor{blue}{Condition \eqref{eq:cond_local_isometry} follows from \cite[{Lemma 2}]{segarra2016blind} and
\begin{equation}
\label{eq:local_isometry_sparse_explicit_linked}
\begin{split}
\|\project_{\mT_\sourceIndex}\decoupledmodelop_\sourceIndex^*\decoupledmodelop_\sourceIndex\project_{\mT_\sourceIndex} - \project_{\mT_\sourceIndex}\| \le \|{\modelop_\sourceIndex^*}_{\mS_\sourceIndex} {\modelop_\sourceIndex}_{\mS_\sourceIndex} - \bI_{\mS_\sourceIndex}\|.
\end{split}
\end{equation}}
%
%
\hspace{-2mm} \textcolor{blue}{Condition \eqref{eq:cond_operator_norm} from $\|\modelop_\sourceIndex\| \le \gamma$ \cite[{Lemma 3}]{segarra2016blind} and the fact that $\norm{\decoupledmodelop_{\sourceIndex}} = \norm{{\modelop_{\sourceIndex}}_{\mS_\sourceIndex}} \le \norm{\modelop_{\sourceIndex}}$.}
The probabilities with which these two conditions hold are revisited in \Cref{sec:union_bound}.
In the following two sections we develop proofs for the remaining three conditions: the coherence between pairs of linear operators~\eqref{eq:cond_cross_coherence} is shown in \Cref{sec:coherence_pairs} and the two dual certificate conditions \eqref{eq:cond_dual0_sparse} and \eqref{eq:cond_dual1_sparse} are shown in \Cref{sec:dual_certificate}. 
Finally, \Cref{sec:union_bound} synthesizes the probabilities to fulfill each of the five conditions.

\subsubsection{Coherence between pairs of linear operators}
\label{sec:coherence_pairs}

A measure of the \emph{coherence} between \textcolor{blue}{mixing components} is $\mu$ as defined in \eqref{eq:cond_cross_coherence}. 
The following lemma provides a lower bound on the probability that~\eqref{eq:cond_cross_coherence} holds.
\begin{mylemma}\label{lemma:coherence}
	Condition~\eqref{eq:cond_cross_coherence} holds with probability at least
	\begin{align}\label{eq:cross_coherence_lemma_prob}
		1 - 2\numNodes\textcolor{blue}{\frac{\numNodes}{\partionSize}} \textcolor{blue}{\min_{\otherSourceIndex\neq\yetOtherSourceIndex}} \exp \! \Bigg(\!-\!\frac{1}{32\numSources^2} \!
		\Bigg[ &\frac{16 \numNodes^2}{9\partionSize^2} \rho_{\bPsi_\yetOtherSourceIndex}^{(\numTaps_\yetOtherSourceIndex)} \! \rho_{\bPsi_\otherSourceIndex}^{(\numTaps_\otherSourceIndex)} \max \! \left\{\! \rho_{\bU_\yetOtherSourceIndex}^{(\numNonzero_\yetOtherSourceIndex)} \!, \rho_{\bU_\otherSourceIndex}^{(\numNonzero_\otherSourceIndex)} \! \right\}  \nonumber \\
		+ &\frac{1}{9\numSources} \frac{\numNodes}{\partionSize} \kappa_{\bU_{\yetOtherSourceIndex}, \bU_{\otherSourceIndex}}^{(\numNonzero_\yetOtherSourceIndex,\numNonzero_\otherSourceIndex)} \kappa_{\bPsi_{\yetOtherSourceIndex}, \bPsi_{\otherSourceIndex}}^{(\numTaps_\yetOtherSourceIndex,\numTaps_\otherSourceIndex)}\Bigg]^{-1}\Bigg).
	\end{align}
\end{mylemma}

\begin{myproof}
\textcolor{blue}{Since the forthcoming analysis does not require explicit knowledge of $\bC_{\mS_\sourceIndex}\transpose$ (cf. \textbf{Assumption 2}, \Cref{S:Recovery}), we shall show that
\begin{equation}
\label{eq:mutual_coherence_sparse}
\max_{\otherSourceIndex \neq \yetOtherSourceIndex} \|{\modelop_\otherSourceIndex^*}_{\mS_\otherSourceIndex} {\modelop_\yetOtherSourceIndex}_{\mS_\yetOtherSourceIndex}\| \le \frac{1}{4\numSources}
\end{equation}
and, similarly to~\eqref{eq:local_isometry_sparse_explicit_linked}, leverage that
\begin{equation}
\label{eq:mutual_coherence_sparse_explicit_linked}
\|\project_{\mT_\otherSourceIndex}\decoupledmodelop_\otherSourceIndex^*\decoupledmodelop_\yetOtherSourceIndex\project_{\mT_\yetOtherSourceIndex}\| \le \|{\modelop_\otherSourceIndex^*}_{\mS_\otherSourceIndex} {\modelop_\yetOtherSourceIndex}_{\mS_\yetOtherSourceIndex}\|.
\end{equation}}
%

Consider the following expression, more general than \textcolor{blue}{the argument of the $\max$ in \eqref{eq:mutual_coherence_sparse}},
\begin{equation}
\label{general_sparsity}
\|{\modelop_{\otherSourceIndex,\partionIndex}^*}_{\mS_\otherSourceIndex} {\modelop_{\yetOtherSourceIndex,\partionIndex}}_{\mS_\yetOtherSourceIndex} (\bZ \parimat_{\yetOtherSourceIndex,\partionIndex})\|.
\end{equation}
To see that \eqref{general_sparsity} indeed generalizes \textcolor{blue}{the term in \eqref{eq:mutual_coherence_sparse}}, consider a partition given by a single set $\partion_1$. 
In this case, from $\bPsi_{\sourceIndex}\hermitian \bPsi_{\sourceIndex} = \bI_{\numTaps_\sourceIndex}$~\footnote{{Under the assumption that the \emph{known} GSOs have distinct eigenvalues [cf. ({AS1}) in \Cref{Ss:Scope}], it can be assumed without loss of generality that $\bPsi_{\sourceIndex}\hermitian \bPsi_{\sourceIndex} = \bI_{\numTaps_\sourceIndex}$. First of all, since each $\bS_\sourceIndex$ has distinct eigenvalues, then the matrices $\bPsi_\sourceIndex$ are full-rank. Second, consider the SVD of $\bPsi_\sourceIndex=\bL_\sourceIndex \bSigma_\sourceIndex \bR_\sourceIndex\hermitian$; accordingly the frequency response of the filter $\bPsi_\sourceIndex\bh_\sourceIndex=\bL_\sourceIndex\bSigma_\sourceIndex\bR_\sourceIndex\hermitian\bh_\sourceIndex$, and define ${\bh'}_\sourceIndex:=\bSigma_\sourceIndex\bR_\sourceIndex\bh_\sourceIndex$. Since $\bPsi_\sourceIndex$ is full-rank, $\bh_\sourceIndex$ can be recovered from ${\bh'}_\sourceIndex$, and finally since the matrix of left-singular vectors matrix holds that $\bL_\sourceIndex\hermitian \bL_\sourceIndex=\bI_{\numTaps_\sourceIndex}$, we can assume that $\bPsi_{\sourceIndex}\hermitian \bPsi_{\sourceIndex} = \bI_{\numTaps_\sourceIndex}$.}} it follows that $\partmat_{\sourceIndex,1} = \parimat_{\sourceIndex,1} = \bI_{\numTaps_\sourceIndex}$. Also, since $\partion_1 = \{1,2,\dots,\numNodes\}$ then $\modelop_{\sourceIndex,1} = \modelop_{\sourceIndex}$ in the single-set partition case.

We shall proceed setting $\otherSourceIndex=2$ and $\yetOtherSourceIndex=1$ to simplify the notation and work with the expression divided into two parts,
\begin{equation*}
{\modelop_{2,\partionIndex}^*}_{\mS_2}(\bz) = \sum_{\ell \in \partion_\partionIndex} z_\ell {\overline{\bu_{2,\ell}}}_{\mS_2} \bpsi_{2,\ell}\hermitian,
\end{equation*}
\begin{equation*}
{\modelop_{1,\partionIndex}}_{\mS_1}(\bZ \parimat_{1,\partionIndex}) =  \left\{ {\bu_{1,\ell}\transpose}_{\Omega_1} \bZ \parimat_{1,\partionIndex} \bpsi_{1,\ell} \right\}_{\ell \in \partion_\partionIndex}.
\end{equation*}
Concatenating the two operators above,
\begin{equation*}
\sum_{\ell \in \partion_\partionIndex}({\bu_{1,\ell}\transpose}_{\Omega_1} \bZ \parimat_{1,\partionIndex}\bpsi_{1,\ell}) \overline{\bu_{2,\ell}}_{\Omega_2} \bpsi_{2,\ell}\hermitian
\end{equation*}
and examining each summand we have that
\begin{equation*}
\left({\bu_{1,\ell}\transpose}_{\Omega_1} \bZ \parimat_{1,\partionIndex}\bpsi_{1,\ell}\right) \overline{\bu_{2,\ell}}_{\Omega_2} \bpsi_{2,\ell}\hermitian =  \overline{\bu_{2,\ell}}_{\Omega_2} {\bu_{1,\ell}\transpose}_{\Omega_1} \bZ \parimat_{1,\partionIndex} \bpsi_{1,\ell} \bpsi_{2,\ell}\hermitian.
\end{equation*}
We seek to bound~\eqref{general_sparsity} by applying the non-commutative Bernstein inequality for matrices \cite[{Theorem 2}]{segarra2016blind} to a sum of matrices given by $\overline{\bu_{2,\ell}}_{\Omega_2} {\bu_{1,\ell}\transpose}_{\Omega_1} \bZ \ \parimat_{1,\partionIndex} \bpsi_{1,\ell} \bpsi_{2,\ell}\hermitian$.
Following the notation in~\cite[{Theorem 2}]{segarra2016blind}, the operators to bound in the concentration inequality are a sequence of
\begin{equation*}
\mZ_\ell(\bZ) = \overline{\bu_{2,\ell}}_{\Omega_2} {\bu_{1,\ell}\transpose}_{\Omega_1} \bZ \ \parimat_{1,\partionIndex} \bpsi_{1,\ell} \bpsi_{2,\ell}\hermitian.
\end{equation*}
Note that, given its definition in~\eqref{eq:partion_matrices}, $\parimat_{1,\partionIndex}$ is Hermitian. 
Then, the adjoint operators become
\begin{equation*}
	\mZ_\ell^*(\bZ) = \overline{\bu_{1,\ell}}_{\Omega_1} {\bu_{2,\ell}\transpose}_{\Omega_2} \bZ \ \bpsi_{2,\ell} \bpsi_{1,\ell}\hermitian \parimat_{1,\partionIndex}.
\end{equation*}
Recalling that $\numNonzero_\sourceIndex=|\Omega_\sourceIndex|$ and leveraging the definition of $\kappa$ in \eqref{eq:kappa}, we have that
\begin{equation}
	\label{eq:coherence_concentration_inequality_center}
		\|\mathcal{Z}_\ell\| \le \|\parimat_{1,\partionIndex}\| \kappa_{\bU_{1}, \bU_{2}}^{\left(\numNonzero_1,\numNonzero_2\right)} \kappa_{\bPsi_{1}, \bPsi_{2}}^{\left(\numTaps_1,\numTaps_2\right)}
		\le \frac{4\numNodes}{3\partionSize} \kappa_{\bU_{1}, \bU_{2}}^{\left(\numNonzero_1,\numNonzero_2\right)} \kappa_{\bPsi_{1}, \bPsi_{2}}^{\left(\numTaps_1,\numTaps_2\right)}
\end{equation}
where $\|\parimat_{\sourceIndex,\partionIndex}\|\!\le\!4\numNodes/(3\partionSize) \ \forall \ (\textcolor{blue}{\sourceIndex},\partionIndex)\!\in\!\lbrace 1,2,\dots,\numSources \rbrace \times \lbrace 1,2,\dots,\numPartionElems \rbrace$ from \cite[(12)]{ling2017blind}. This is a direct consequence of the definitions of the partition $\partion_\partionIndex$ and the matrices $\partmat_{\sourceIndex,\partionIndex}$ lying close to $\bI_{\numTaps_\sourceIndex}$, see to \cite[section II-C]{ling2017blind} for further discussion.

In order to implement the concentration inequality, we compute the expected value of the concatenation operators $\mZ_\ell\mZ_\ell^*(\bZ)$ and $\mZ_\ell^*\mZ_\ell(\bZ)$ to obtain
\begin{align*}
		\ev{\mZ_\ell\mZ_\ell^*(\bZ)} = & \left(\bpsi_{1,\ell}\hermitian \parimat_{1,\partionIndex} \parimat_{1,\partionIndex} \bpsi_{1,\ell}\right) \ev{{\bu_{1,\ell}\transpose}_{\Omega_1} \overline{\bu_{1,\ell}}_{\Omega_1}} \\
		&\ev{\overline{\bu_{2,\ell}}_{\Omega_2} {\bu_{2,\ell}\transpose}_{\Omega_2}} \bZ \ \bpsi_{2,\ell} \bpsi_{2,\ell}\hermitian,
\end{align*}
\begin{align*}
		\ev{\mZ_\ell^*\mZ_\ell(\bZ)} = &\left(\bpsi_{2,\ell}\hermitian \bpsi_{2,\ell}\right) \ev{{\bu_{2,\ell}\transpose}_{\Omega_2} \overline{\bu_{2,\ell}}_{\Omega_2}} \\ & \ev{\overline{\bu_{1,\ell}}_{\Omega_1} {\bu_{1,\ell}\transpose}_{\Omega_1}} \bZ \ \parimat_{1,\partionIndex} \bpsi_{1,\ell} \bpsi_{1,\ell}\hermitian \parimat_{1,\partionIndex}.
\end{align*}
Due to the normalization $\bU_\sourceIndex\hermitian \bU_\sourceIndex = \numNodes \bI_{\numNodes}$, it follows that $\mathbb{E}(\overline{\bu_{\sourceIndex,\ell}}_{\Omega_\sourceIndex} {\bu_{\sourceIndex,\ell}\transpose}_{\Omega_\sourceIndex}) = \bI_{\numNodes_{\Omega_\sourceIndex}}$. 
Moreover, from the definition in~\eqref{eq:rho}, we have that the norms of $\mathbb{E}({\bu_{\sourceIndex,\ell}\transpose}_{\Omega_\sourceIndex} \overline{\bu_{\sourceIndex,\ell}}_{\Omega_\sourceIndex})$ are bounded by $\rho_{\bU_\sourceIndex}^{(\numNonzero_\sourceIndex)}$. 
Therefore,
\begin{align*}
		\left\|\sum_{\ell \in \partion_\partionIndex} \ev{\mZ_\ell\mZ_\ell^*}\right\| &\le 
		\rho_{\bU_1}^{(\numNonzero_1)} \sum_{\ell \in \partion_\partionIndex} \|\bpsi_{1,\ell}\hermitian \parimat_{1,\partionIndex} \parimat_{1,\partionIndex} \bpsi_{1,\ell}\| \ \|\bpsi_{2,\ell} \bpsi_{2,\ell}\hermitian\| \\
		&\le \frac{16 \numNodes^2}{9 \partionSize^2} \rho_{\bU_1}^{(\numNonzero_1)} \rho_{\bPsi_1}^{(\numTaps_1)} \rho_{\bPsi_2}^{(\numTaps_2)},
\end{align*}
\begin{align*}
		\left\|\sum_{\ell \in \partion_\partionIndex} \ev{\mZ_\ell^*\mZ_\ell}\right\| &\le
		\rho_{\bU_2}^{(\numNonzero_2)} \sum_{\ell \in \partion_\partionIndex} \|\bpsi_{2,\ell}\hermitian \bpsi_{2,\ell}\| \ \|\parimat_{1,\partionIndex} \bpsi_{1,\ell} \bpsi_{1,\ell}\hermitian \parimat_{1,\partionIndex}\| \\
		&\le \frac{16 \numNodes^2}{9 \partionSize^2} \rho_{\bU_2}^{(\numNonzero_2)} \rho_{\bPsi_1}^{(\numTaps_1)} \rho_{\bPsi_2}^{(\numTaps_2)}
\end{align*}
where we have used, once again, that $\|\parimat_{\sourceIndex,\partionIndex}\| \le 4\numNodes/(3\partionSize)$. 
Finally, the variance parameter in the Bernstein inequality is given by
\begin{equation}
	\label{eq:coherence_concentration_inequality_variance}
	\sigma^2= \frac{16 \numNodes^2}{9 \partionSize^2} \rho_{\bPsi_1}^{(\numTaps_1)} \rho_{\bPsi_2}^{(\numTaps_2)} \max\left\{\rho_{\bU_1}^{(\numNonzero_1)}, \rho_{\bU_2}^{(\numNonzero_2)}\right\}.
\end{equation}
Employing the concentration inequality~\cite[{Theorem 2}]{segarra2016blind} with the parameters obtained in \eqref{eq:coherence_concentration_inequality_center} and \eqref{eq:coherence_concentration_inequality_variance}, \textcolor{blue}{taking the \emph{union bound} over $1\le\partionIndex\le\numPartionElems$ (recall that $\numNodes=\partionSize\numPartionElems$), and introducing $\min_{1\le\otherSourceIndex\neq\yetOtherSourceIndex\le\numSources}$ (to arrive at the left-hand side in \eqref{eq:mutual_coherence_sparse} from $\mZ_\ell$ with fixed $\otherSourceIndex,\yetOtherSourceIndex,\partionIndex$), and \eqref{eq:mutual_coherence_sparse_explicit_linked}} we have that
\begin{align*}
&\bP\!\left(\mu \!\ge \!t\right) \! \le \! 2\numNodes \textcolor{blue}{\frac{\numNodes}{\partionSize} \min_{\otherSourceIndex\neq\yetOtherSourceIndex}} \exp\Bigg(\!\!-\frac{t^2}{2} \\
&\Bigg[\frac{16 \numNodes^2}{9 \partionSize^2} \!\rho_{\bPsi_\yetOtherSourceIndex}^{(\numTaps_\yetOtherSourceIndex)}\! \rho_{\bPsi_\otherSourceIndex}^{(\numTaps_\otherSourceIndex)} \!\max\left\{\!\rho_{\bU_\yetOtherSourceIndex}^{(\numNonzero_\yetOtherSourceIndex)}\!, \rho_{\bU_\otherSourceIndex}^{(\numNonzero_\otherSourceIndex)}\!\right\}\!+\!\frac{t}{3} \frac{4\numNodes}{3\partionSize} \kappa_{\bU_{\yetOtherSourceIndex}, \bU_{\otherSourceIndex}}^{\left(\numNonzero_\yetOtherSourceIndex,\numNonzero_\otherSourceIndex\right)} \kappa_{\bPsi_{\yetOtherSourceIndex}, \bPsi_{\otherSourceIndex}}^{\left(\numTaps_\yetOtherSourceIndex,\numTaps_\otherSourceIndex\right)}\Bigg]^{\!-1}\Bigg)\!.
\end{align*}
The final result of the lemma is obtained by setting $t=1/(4 \numSources)$ and switching the lower bound to the upper bound.
\end{myproof}

\subsubsection{Dual certificate}
\label{sec:dual_certificate}

The approximate dual certificate $\{\bY_\sourceIndex\}_{\sourceIndex=1}^\numSources$ is constructed via the golfing scheme, producing a sequence of random matrices \cite{ling2017blind}. The purpose behind the method is that the sequence $\{\bY_{\sourceIndex,\partionIndex}\}_{\partionIndex=1}^\numPartionElems$ approaches $\textcolor{blue}{\bs}_\sourceIndex^{(*)}\bh_\sourceIndex^{(*)}{\transpose}$ on $\textcolor{blue}{\mT}_\sourceIndex$ exponentially fast while keeping $\bY_{\sourceIndex,\partionIndex}$ small on $\textcolor{blue}{\mT}^\perp_\sourceIndex$. 
The method is now outlined.
\begin{enumerate}
	\item Initialization:
	\begin{align}
		\blambda_0 &:= \sum_{\otherSourceIndex=1}^\numSources \textcolor{blue}{\decoupledmodelop}_{\otherSourceIndex,\textcolor{blue}{1}}(\textcolor{blue}{\bs}_\otherSourceIndex^{(*)}\bh_\otherSourceIndex^{(*)}{\transpose} \parimat_{\otherSourceIndex,\textcolor{blue}{1}}) \ \in \mathbb{C}^\numNodes, \\
		\bY_{\sourceIndex,0} &:= \bzero_{\textcolor{blue}{\numNonzero_\sourceIndex} \times \numTaps_\sourceIndex}, \quad \forall \ \sourceIndex \in \{1,2,\dots,\numSources\}.
	\end{align}
	
	\item Recursion:
	\begin{align}
		\blambda_{\partionIndex-1} &:= \sum_{\otherSourceIndex=1}^\numSources \textcolor{blue}{\decoupledmodelop}_{\otherSourceIndex,\partionIndex}\left((\textcolor{blue}{\bs}_\otherSourceIndex^{(*)}\bh_\otherSourceIndex^{(*)}{\transpose} - \textcolor{blue}{\project_{\mT_\otherSourceIndex}}(\bY_{\otherSourceIndex,\partionIndex-1})) \parimat_{\otherSourceIndex,\partionIndex}\right), \label{eq:golfing_recursion} \\
		\bY_{\sourceIndex,\partionIndex} &:= \bY_{\sourceIndex,\partionIndex-1} + \textcolor{blue}{\decoupledmodelop}^*_{\sourceIndex,\partionIndex}(\blambda_{\partionIndex-1}),  \quad \forall \ \sourceIndex \in \{1,2,\dots,\numSources\}. \label{eq:golfing_recursion_2}
	\end{align}
\end{enumerate}
Let $\bY_{\sourceIndex} := \bY_{\sourceIndex,\numPartionElems}$ for all $\sourceIndex$ and $\blambda:=\blambda_{\numPartionElems}$. Furthermore, the difference (residuals)
\begin{equation}
	\label{eq:cond_dual1_residual}
	\bW_{\sourceIndex,\partionIndex} := \textcolor{blue}{\bs}_\sourceIndex^{(*)}\bh_\sourceIndex^{(*)}{\transpose} - \textcolor{blue}{\project_{\mT_\sourceIndex}}(\bY_{\sourceIndex,\partionIndex}), \quad \bW_{\sourceIndex,0} := \textcolor{blue}{\bs}_\sourceIndex^{(*)}\bh_\sourceIndex^{(*)}{\transpose}
\end{equation}
leads to rewriting
\begin{equation}
	\blambda_{\partionIndex-1} = \sum_{\otherSourceIndex=1}^\numSources \textcolor{blue}{\decoupledmodelop}_{\otherSourceIndex,\partionIndex}(\bW_{\otherSourceIndex,\partionIndex-1}\parimat_{\otherSourceIndex,\partionIndex})
\end{equation}
and the recursion
\begin{equation}
	\label{eq:golf_diff_recursion}
	\bW_{\sourceIndex,\partionIndex} = \bW_{\sourceIndex,\partionIndex-1} - \sum_{\otherSourceIndex=1}^\numSources \textcolor{blue}{\project_{\mT_\sourceIndex}} {\textcolor{blue}{\decoupledmodelop}^*_{\textcolor{blue}{\sourceIndex},\partionIndex}} {\textcolor{blue}{\decoupledmodelop}_{\otherSourceIndex,\partionIndex}}(\bW_{\otherSourceIndex,\partionIndex-1} \parimat_{\otherSourceIndex,\partionIndex}).
\end{equation}
\textcolor{blue}{Note that, for the residuals, it holds that $\project_{\mT_\otherSourceIndex^\perp}(\bW_{\otherSourceIndex,\partionIndex-1})=0$}.

{The dual certificate comprises two conditions: the exponential decay of $\|\textcolor{blue}{\bs}_\sourceIndex^{(*)}\bh_\sourceIndex^{(*)}{\transpose}- \textcolor{blue}{\project_{\mT_\sourceIndex}}(\bY_{\sourceIndex})  \|_{\text{F}}$, given in {Lemma}~\autoref{lemma:first_dual_certificate}, and $\|\textcolor{blue}{\project_{\mT_\sourceIndex^\perp}}(\bY_{\sourceIndex})\| \le 1/2$ in {Lemma}~\autoref{lemma:second_dual_certificate}.}

\begin{mylemma}\label{lemma:first_dual_certificate} Under the condition that $\numPartionElems\!\ge\!\log_2(5\numSources\gamma)$, then $\|\textcolor{blue}{\bs}_\sourceIndex^{(*)}\bh_\sourceIndex^{(*)}{\transpose}- \textcolor{blue}{\project_{\mT_\sourceIndex}}(\bY_{\sourceIndex})\|_{\text{F}} \le (5 \numSources \gamma)^{-1}$ with probability at least $1-\numNodes^{-\alpha+1}$.\end{mylemma}

\begin{myproof}
The following development follows the steps from \cite[{Lemma 5.12}]{ling2017blind}. Expanding the sum in \eqref{eq:golf_diff_recursion},
\begin{equation}
	\begin{split}
		\bW_{\sourceIndex,\partionIndex} = \bW_{\sourceIndex,\partionIndex-1} - &\textcolor{blue}{\project_{\mT_\sourceIndex}}\textcolor{blue}{\decoupledmodelop}^*_{\sourceIndex,\partionIndex} \textcolor{blue}{\decoupledmodelop}_{\sourceIndex,\partionIndex} (\bW_{\sourceIndex,\partionIndex-1} \parimat_{\sourceIndex,\partionIndex}) \\ - \sum_{\otherSourceIndex=1, \ \otherSourceIndex \neq \sourceIndex}^\numSources &\textcolor{blue}{\project_{\mT_\sourceIndex}}{\textcolor{blue}{\decoupledmodelop}^*_{\textcolor{blue}{\sourceIndex},\partionIndex}} {\textcolor{blue}{\decoupledmodelop}_{\otherSourceIndex,\partionIndex}} (\bW_{\otherSourceIndex,\partionIndex-1} \parimat_{\otherSourceIndex,\partionIndex}).
	\end{split}
\end{equation}
Applying \cite[{Lemma 2}]{segarra2016blind} ($\delta=1/4$, $\numPartionElems=1$ therein), \textcolor{blue}{\eqref{eq:local_isometry_sparse_explicit_linked}}, $\mu \le 1/(4\numSources)$ (cf. {Lemma~\ref{lemma:coherence}}), and noting that $\bW_{\otherSourceIndex,\partionIndex-1} \textcolor{blue}{\in \mT_\otherSourceIndex}$:
\begin{equation}
	\|\bW_{\sourceIndex,\partionIndex}\|_{\text{F}} \le \frac{1}{4} \|\bW_{\sourceIndex,\partionIndex-1}\|_{\text{F}} + \frac{1}{4 \numSources} \sum_{\otherSourceIndex\neq\sourceIndex} \|\bW_{\otherSourceIndex,\partionIndex-1}\|_{\text{F}}, \quad 1 \le \sourceIndex \le \numSources
\end{equation}
and simplifying with the max
\begin{equation}
	\|\bW_{\sourceIndex,\partionIndex}\|_{\text{F}} \le \frac{1}{2} \max_{1 \le \sourceIndex \le \numSources} \|\bW_{\sourceIndex,\partionIndex-1}\|_{\text{F}}, \quad 1 \le \sourceIndex \le \numSources.
	\label{eq:dc_cond1}
\end{equation}
Finally, recalling that $\|\bW_{\sourceIndex,0}\|_{\text{F}}=\|\textcolor{blue}{\bs}_\sourceIndex^{(*)}\bh_\sourceIndex^{(*)}{\transpose}\|_{\text{F}}=1$, it is proven that $\|\textcolor{blue}{\bs}_\sourceIndex^{(*)}\bh_\sourceIndex^{(*)}{\transpose} - \textcolor{blue}{\project_{\mT_\sourceIndex}}(\bY_{\sourceIndex})\|_{\text{F}}$ decays exponentially because by induction
\begin{equation}
	\label{eq:induction_dual_certificate}
	\|\bW_{\sourceIndex,\partionIndex}\|_{\text{F}} \le 2^{-\partionIndex}, \quad 1 \le \partionIndex \le \numPartionElems, \quad 1 \le \sourceIndex \le \numSources.
\end{equation}
{Equating the most constraining upper bound from \eqref{eq:induction_dual_certificate} $2^{-\numPartionElems}$ to the upper bound sought $(5 \numSources \gamma)^{-1}$~\eqref{eq:cond_dual0_sparse}, a lower bound on the number of partition elements is obtained, $\numPartionElems\!\ge\!\log_2(5\numSources\!\gamma)$. Such partitioning guarantees the fulfillment of $\|\textcolor{blue}{\bs}_\sourceIndex^{(*)}\bh_\sourceIndex^{(*)}{\transpose}- \textcolor{blue}{\project_{\mT_\sourceIndex}}(\bY_{\sourceIndex})\|_{\text{F}} \le (5 \numSources \gamma)^{-1}$ (cf. \cite[section II-C]{ling2017blind} for further discussion on the partitioning)}.
\end{myproof}

\begin{mylemma}\label{lemma:second_dual_certificate} Under the condition that \newline $\boundDualCertificateSecondCondition$, then $\|\textcolor{blue}{\project_{\mT_\sourceIndex^\perp}({\bY_{\sourceIndex}})}\| \le 1/2$ with probability at least $1-\numNodes^{-\alpha+1}$.\end{mylemma}

\begin{myproof}
	Expanding the recursion in \eqref{eq:golfing_recursion}-\eqref{eq:golfing_recursion_2},
	\begin{equation}
		\bY_{\sourceIndex} = \sum_{\partionIndex=1}^\numPartionElems \textcolor{blue}{\decoupledmodelop}_{\sourceIndex,\partionIndex}^* \left( \blambda_{\partionIndex-1} \right).
	\end{equation}
	Since \textcolor{blue}{$\project_{\mT^\perp_\sourceIndex}(\bW_{\sourceIndex,\partionIndex-1})=0$} (cf. comment after \eqref{eq:golf_diff_recursion}), it holds that
	%
	%
	\textcolor{blue}{
	\begin{equation}
		\begin{split}
			\| \project_{\mT^\perp_\sourceIndex}(\bY_{\sourceIndex}) \| &= \left\| \project_{\mT^\perp_\sourceIndex}\left(\sum_{\partionIndex=1}^\numPartionElems \decoupledmodelop_{\sourceIndex,\partionIndex}^*(\blambda_{\partionIndex-1}) -  {\bW_{\sourceIndex,\partionIndex-1}}\right) \right\| \\ &\le \sum_{\partionIndex=1}^\numPartionElems \| \decoupledmodelop_{\sourceIndex,\partionIndex}^* \left( \blambda_{\partionIndex-1} \right) - \bW_{\sourceIndex,\partionIndex-1} \|.
		\end{split}
	\end{equation}}
	From here, if we show that $\| \decoupledmodelop_{\sourceIndex,\partionIndex}^* \left( \blambda_{\partionIndex-1} \right) - \bW_{\sourceIndex,\partionIndex-1} \| \le 2^{-\partionIndex-1}$, then $\| \textcolor{blue}{\project_{\mT^\perp_\sourceIndex}(\bY_{\sourceIndex})} \| \le \sum_{\partionIndex=1}^\numPartionElems 2^{-\partionIndex-1} \le 1/2$ follows directly.
	
	The upper bound $\|\decoupledmodelop_{\sourceIndex,\partionIndex}^*( \blambda_{\partionIndex-1}) - \bW_{\sourceIndex,\partionIndex-1}\| \le 2^{-\partionIndex-1}$ is based on \cite[{Lemma 5.14}]{ling2017blind}. \textcolor{blue}{Nevertheless, the $\bA_\sourceIndex$ matrices in \cite{ling2017blind} are Gaussian and therefore the resulting condition on $\partionSize$ is slightly different. Let us now illustrate this difference, that is, the results in \cite[{Lemma 5.14}]{ling2017blind} adapted to (projected) random unitary matrices.}
	
	Firstly, the random matrices in the concentration inequality \cite[(56)]{ling2017blind} are bounded by
	%
	\begin{equation}
		\begin{split}
			\label{eq:dual_second_proof_Zl}
			\|\mZ_\ell\| &\le \frac{\mu_{\max} \sqrt{\numTaps_\sourceIndex}}{\sqrt{\numNodes}} \textcolor{blue}{\left(\!(\rho_{\bU_\sourceIndex}^{(\numNonzero_\sourceIndex)}\!+\!1) \! + \! \sum_{\substack{\otherSourceIndex=1\\ \otherSourceIndex \neq \sourceIndex}}^{\numSources}{\!\sqrt{\rho_{\bU_{\otherSourceIndex}}^{(\numNonzero_\otherSourceIndex)} \rho_{\bU_{\sourceIndex}}^{(\numNonzero_\sourceIndex)}}}\!\right)} \frac{\sqrt{\numNodes}}{\partionSize}  2^{-\partionIndex+1}  \! \mu_h \\
		\end{split}
	\end{equation}
	The remaining steps follow the development in \cite[{Lemma 5.14}]{ling2017blind}, we only need to keep track of the additional factors stemming from the random unitary model. \textcolor{blue}{In fact, accounting for these concentration factor across all $\bU_\sourceIndex$, let us define}
	\begin{equation}
		\textcolor{blue}{\rho_\text{max} := \max_{1 \le \sourceIndex \le \numSources} \rho_{\bU_\sourceIndex}^{(\numNonzero_\sourceIndex)}}
	\end{equation}
	The center and standard deviation parameters become, respectively,
	\begin{equation}
		\label{eq:dual_second_proof_R}
		\max_{\ell \in \partion_\partionIndex} \| \mZ_\ell \| \le \frac{\numSources}{\partionSize} 2^{-\partionIndex+1} \max\lbrace \mu_{\max}^2 \numTaps_\sourceIndex,\mu_h^2 \textcolor{blue}{(\rho_\text{max}+1)^2}\rbrace
	\end{equation}
	%
	%
	\begin{equation}
		\label{eq:dual_second_proof_sigma}
		\sigma^2 \le \mathcal{C} \frac{\numSources}{\partionSize} 4^{-\partionIndex+1}  \textcolor{blue}{\max\{\mu_{\max}^2 \numTaps_\sourceIndex \rho_\text{max}, \mu_h^2 \numNonzero_\sourceIndex\}}
	\end{equation}
	Having the vectors \textcolor{blue}{$\ba_{\sourceIndex,\nodeIndex}$} random unitary instead of Gaussian \textcolor{blue}{and $\bq \in \mathbb{C}^{\numNonzero_\sourceIndex}$}, to obtain \eqref{eq:dual_second_proof_Zl} and thus \eqref{eq:dual_second_proof_R}, \cite[(90)]{ling2017blind} becomes
	\begin{equation}
		\textcolor{blue}{\|(\ba_{\sourceIndex,\ell} \ba_{\sourceIndex,\ell}\hermitian  - \bI_{\numNonzero_\sourceIndex}) \bq \| \le |\ba_{\sourceIndex,\ell}\hermitian \bq| \|\ba_{\sourceIndex,\ell}\|+ \| \bq \| \le (\rho_\text{max} + 1) \| \bq \|}.
	\end{equation}
	In addition, to obtain \eqref{eq:dual_second_proof_sigma}, \cite[(91)]{ling2017blind} becomes
	%
	%
	\begin{equation}
		\textcolor{blue}{\left\|\mathbb{E}\left[|\bq\hermitian \ba_{\sourceIndex,\ell}|^2 \ba_{\sourceIndex,\ell} \ba_{\sourceIndex,\ell}\hermitian\right]\right\| \le \|\bq\|^2 \rho_{\bU_{\sourceIndex}}^{(\numNonzero_\sourceIndex)}}.
    \end{equation}
	\textcolor{blue}{since $|\bq\hermitian \ba_{\sourceIndex,\ell}|\!\le\! \|\bq\|\sqrt{\rho_{\bU_{\sourceIndex}}^{(\numNonzero_\sourceIndex)}}$ and $\mathbb{E}[\ba_{\sourceIndex,\ell} \ba_{\sourceIndex,\ell}\hermitian]\!=\!\bI_{\numNonzero_\sourceIndex}$}.
	Finally, we let $\boundDualCertificateSecondCondition$ so that $\|\decoupledmodelop_{\sourceIndex,\partionIndex}^* \left( \blambda_{\partionIndex-1} \right) - \bW_{\otherSourceIndex,\partionIndex-1} \| \le 2^{-\partionIndex-1}$ with probability at least $1-\numNodes^{-\alpha+1}$ simultaneously $\forall \ \sourceIndex \in \{1,2,\dots,\numSources\}$ and $\partionIndex \in \{1,2,\dots,\numPartionElems\}$.
\end{myproof}

\subsubsection{Union bound: Fusing the sufficient conditions}
\label{sec:union_bound}

The local isometry property \eqref{eq:cond_local_isometry}, \textcolor{blue}{follows from \eqref{eq:local_isometry_sparse_explicit_linked} and} $\|{\modelop_\sourceIndex^*}_{\mS_\sourceIndex} {\modelop_\sourceIndex}_{\mS_\sourceIndex}-\bI_{\mS_\sourceIndex}\|\!\le\!1/4$, which is proven in \cite[{Lemma 2}]{segarra2016blind}. In particular, it is obtained setting $\partionSize=\numNodes$ and $\delta=1/4$ in \cite[(30)]{segarra2016blind} and the upper bound holds with probability at least $1-\numNodes^{-\alpha+1}$ if $\alpha \ge 1$ and
\begin{equation}
	\label{eq:alpha_1}
	\begin{split}
		\alpha \le \alpha_1 := &\min_{1 \le \sourceIndex \le \numSources} \frac{1/16 \ (5/2 + 1/6)^{-1}}{\rho_{\bPsi_\sourceIndex}^{(\numTaps_\sourceIndex)} \rho_{\bU_\sourceIndex}^{(\numNonzero_\sourceIndex)} \log\left(2 \numNodes \numTaps_\sourceIndex \numNonzero_\sourceIndex\right)} = \\ &\min_{1 \le \sourceIndex \le \numSources} \frac{3}{128} \left(\rho_{\bPsi_\sourceIndex}^{(\numTaps_\sourceIndex)} \rho_{\bU_\sourceIndex}^{(\numNonzero_\sourceIndex)} \log\left(2 \numNodes \numTaps_\sourceIndex \numNonzero_\sourceIndex\right)\right)^{-1}.
	\end{split}
\end{equation}
(cf. \cite[(31)]{segarra2016blind}).

The coherence between pairs \eqref{eq:cond_cross_coherence}, $\mu\le1/(4\numSources)$, is proven in \Cref{sec:coherence_pairs}. Setting $\partionSize=\numNodes$ in \eqref{eq:cross_coherence_lemma_prob}, it holds with probability at least $1-\numNodes^{-\alpha+1}$ if $\alpha \ge 1$ and
\begin{equation}
	\begin{split}
		\alpha \le \alpha_2 := &\frac{9}{32\numSources^2\textcolor{blue}{\log\numNodes}} \min_{\otherSourceIndex \neq \yetOtherSourceIndex} \Bigg( \frac{1}{\numSources} \kappa_{\bU_{\yetOtherSourceIndex}, \bU_{\otherSourceIndex}}^{\left(\numNonzero_\yetOtherSourceIndex,\numNonzero_\otherSourceIndex\right)} \kappa_{\bPsi_{\yetOtherSourceIndex}, \bPsi_{\otherSourceIndex}}^{\left(\numTaps_\yetOtherSourceIndex,\numTaps_\otherSourceIndex\right)} + \\ &\rho_{\bPsi_\yetOtherSourceIndex}^{(\numTaps_\yetOtherSourceIndex)} \rho_{\bPsi_\otherSourceIndex}^{(\numTaps_\otherSourceIndex)} \max\left\{\rho_{\bU_\yetOtherSourceIndex}^{(\numNonzero_\yetOtherSourceIndex)}, \rho_{\bU_\otherSourceIndex}^{(\numNonzero_\otherSourceIndex)}\right\}\Bigg)^{-1}.
	\end{split}	
\end{equation}

The third condition \eqref{eq:cond_operator_norm}, \textcolor{blue}{fullfilled by $\|\modelop_\sourceIndex\| \le \gamma$}, which is proven in \cite[{Lemma 3}]{segarra2016blind}, holds with probability at least $1-\numNodes^{-\alpha+1}$, $\gamma=\sqrt{2\numNodes(\log(2\numTaps_\sourceIndex\numNodes)+1)+1}$, $\alpha\ge1$, and
\begin{equation}
	\alpha \le \min_{1 \le \sourceIndex \le \numSources} \left(\rho_{\bPsi_\sourceIndex}^{(\numTaps_\sourceIndex)} \log\numNodes\right)^{-1},
\end{equation}
which is already fulfilled via $\alpha \le \alpha_1$ in~\eqref{eq:alpha_1}.

The first dual certificate condition \eqref{eq:cond_dual0_sparse} is based on two of the previous conditions as shown in \Cref{sec:dual_certificate}, the local isometry property \eqref{eq:cond_local_isometry} and the coherence between pairs \eqref{eq:cond_cross_coherence}.

The second dual certificate and last condition \eqref{eq:cond_dual1_sparse}, holds with probability at least $1-\numNodes^{-\alpha+1}$ and $\boundDualCertificateSecondCondition$ (cf. Lemma~\ref{lemma:second_dual_certificate}). Translating the latter condition into an upper bound on $\alpha$ and simplifying to $\partionSize=\numNodes$,
\begin{equation}
\begin{gathered}
	\alpha \le \alpha_3 := \textcolor{blue}{-\mathcal{C}^{-1}\log\numSources \ +} \\ 
	\textcolor{blue}{\min_{1 \le \sourceIndex \le \numSources} \! \left(\mathcal{C} \numSources (\rho_{\max} \! + \! 1)^2 \! \max \! \left\lbrace\mu_{\max}^2 \numTaps_\sourceIndex, \mu_h^2 \numNonzero_\sourceIndex\right\rbrace \! \log^2\numNodes\right)^{-1} \! \numNodes}
\end{gathered}
\end{equation}
By setting $\alpha \le \min\{\alpha_1, \alpha_2, \alpha_3\}$ as in the statement of the theorem, we guarantee that all conditions \eqref{eq:cond_local_isometry}-\eqref{eq:cond_dual1_sparse} are satisfied with high probability.

\bibliographystyle{IEEEtran}
\balance
\bibliography{citations}

\end{document}